\documentclass[journal]{IEEEtran}
\usepackage{graphicx} 
\usepackage{indentfirst}
\usepackage{algpseudocode}
\usepackage{amsmath}
\usepackage{amsfonts}
\usepackage{amssymb}
\usepackage{multirow}
\usepackage{makecell}
\usepackage{algorithm}
\usepackage[font=small]{caption}
\usepackage{cite}
\usepackage{subcaption}
\usepackage{pifont}
\usepackage{booktabs}
\usepackage{enumitem}
\usepackage{url}

\title{Generalized Query-Oriented Image Semantic Coding Empowered by Large AI Models and Semantic-Aware Hybrid Beamforming}
\author{Sin-Yu Huang, \textit{Graduate Student Member, IEEE}, and Vincent W.S. Wong, \textit{Fellow, IEEE}
\thanks{Manuscript received on Jan. 13, 2026; revised on May 4, 2026 and Jul. 23, 2026; accepted on Jul. 26, 2026. This work was supported in part by the Natural Sciences and Engineering Research Council of Canada and the Digital Research Alliance of Canada (alliancecan.ca). This paper has been published in part in the \textit{Proceedings of IEEE Global Communications Conference (GLOBECOM)}, Dec. 2025 \cite{Huang2025VLM}. The editor coordinating the review of this paper and approving it for publication was Dongfang Xu. (Corresponding author: Vincent W.S. Wong)

S.-Y. Huang and V. W.S. Wong are with the Department of Electrical and Computer Engineering, The University of British Columbia, Vancouver, BC, V6T 1Z4, Canada (e-mail: \{syhuang, vincentw\}@ece.ubc.ca). 

Color versions of one or more of the figures in this paper are available online at https://ieeexplore.ieee.org.
}}

\begin{document}

\maketitle

\begin{abstract}
    Semantic communication is an emerging paradigm that can preserve the meaning of data during transmission. However, human users are often interested in specific semantic content based on their intent, and users' intent is often not considered in current semantic coding design. Moreover, most of the existing semantic models are fine-tuned using specific datasets, which limits their generalization capability. Furthermore, how to prioritize semantically important features in large-scale multiple-input multiple-output orthogonal frequency-division multiplexing (MIMO-OFDM) systems remains largely unexplored. To address the aforementioned challenges, in this paper, we propose a generalized query-oriented image semantic coding (QO-ISC) framework. In the proposed framework, the transmitter extracts features which are relevant to the user's query and the receiver reconstructs an image based on those features. We use a pretrained large artificial intelligence (AI) model (LAM) to enhance general feature representations. We develop a semantic-aware hybrid beamforming (SA-HBF) algorithm to prioritize semantically important features for large-scale MIMO-OFDM system. When evaluated on unseen object categories within the dataset, simulation results show that our proposed generalized QO-ISC framework achieves better performance than the traditional codec and two state-of-the-art semantic coding schemes.
\end{abstract}

\section{Introduction}
\label{Sec:intro}

The advent of intelligent applications, such as autonomous vehicles, augmented reality/virtual reality (AR/VR), and remote surgery, has significantly increased the demand for wireless communication systems capable of delivering high data rates with ultra-reliable and low-latency performance. To  support these applications, the sixth-generation (6G) wireless networks are anticipated to transition from traditional data-oriented communication paradigm to semantic communication paradigm~\cite{Getu2024}. In conventional data-oriented communication, the goal is to accurately reproduce the original message at the receiver with each data bit having equal importance. However, data bits do not uniformly contribute to the conveyed content. Some data bits carry critical semantic meanings, while others are less important. Semantic communication addresses this issue by minimizing the semantic error between the recovered data and the source data, rather than bit-level errors~\cite{Deniz2023}. For example, DeepSC~\cite{Xie2021} is a deep learning-based semantic communication framework that extracts and transmits the semantic features from textual data. Instead of minimizing the mean square error (MSE), DeepSC evaluates the recovered message based on BERTScore~\cite{Zhang2020}, which is  measured by the Bidirectional Encoder Representations from Transformers (BERT)~\cite{Devlin2018}. 
Results showed that DeepSC achieves more reliable transmissions in terms of sentence similarity when compared with traditional coding schemes, such as Huffman coding and turbo coding, particularly under low signal-to-noise ratio (SNR) scenarios.

{Despite the advantages of semantic coding, there are potential challenges that limit its integration in future applications. First, recent semantic communication research focuses on extracting the overall semantic features from the source, rather than prioritizing features that are more relevant to the receiver's intent~\cite{Weng2024, Xu2025, Xiang2025}. In these studies, the general features are extracted from the source data and reconstructed at the receiver. The aforementioned works do not consider the fact that not all semantic components have the same importance from the receiver's perspective. In \cite{Huang2023}, a semantic coding scheme is proposed to jointly optimize the quality of reconstruction and task-oriented performance. In addition to extracting the general features, those features which are related to a specific task are prioritized.
However, the tasks considered in~\cite{Huang2023}, such as detection, segmentation and classification, are all machine-centric and do not fully reflect human-centric communication scenarios, where the intent is often complex and dynamic. For example, consider an image of a horse on a grassy ranch with some kids nearby, as illustrated in Fig.~\ref{fig:scene}. A rancher may ask the question: ``Where is my horse?'', focusing on the animal rather than any humans in the scene. On the other hand, the parents who bring their kids to the ranch may ask the question: ``Are my kids in the farm?'', shifting the attention to the kids. This example illustrates that the importance of semantic content in the ranch scene is highly dependent on the user's intent. This dependency is crucial in human-centric applications such as remote surveillance, where users' intent shift dynamically across different situations. Thus, there is a need for semantic coding frameworks that can dynamically prioritize the content relevant to the user's intent when reconstructing the source data.
 \begin{figure}[t]
     \centering
     \includegraphics[width=0.8\linewidth]{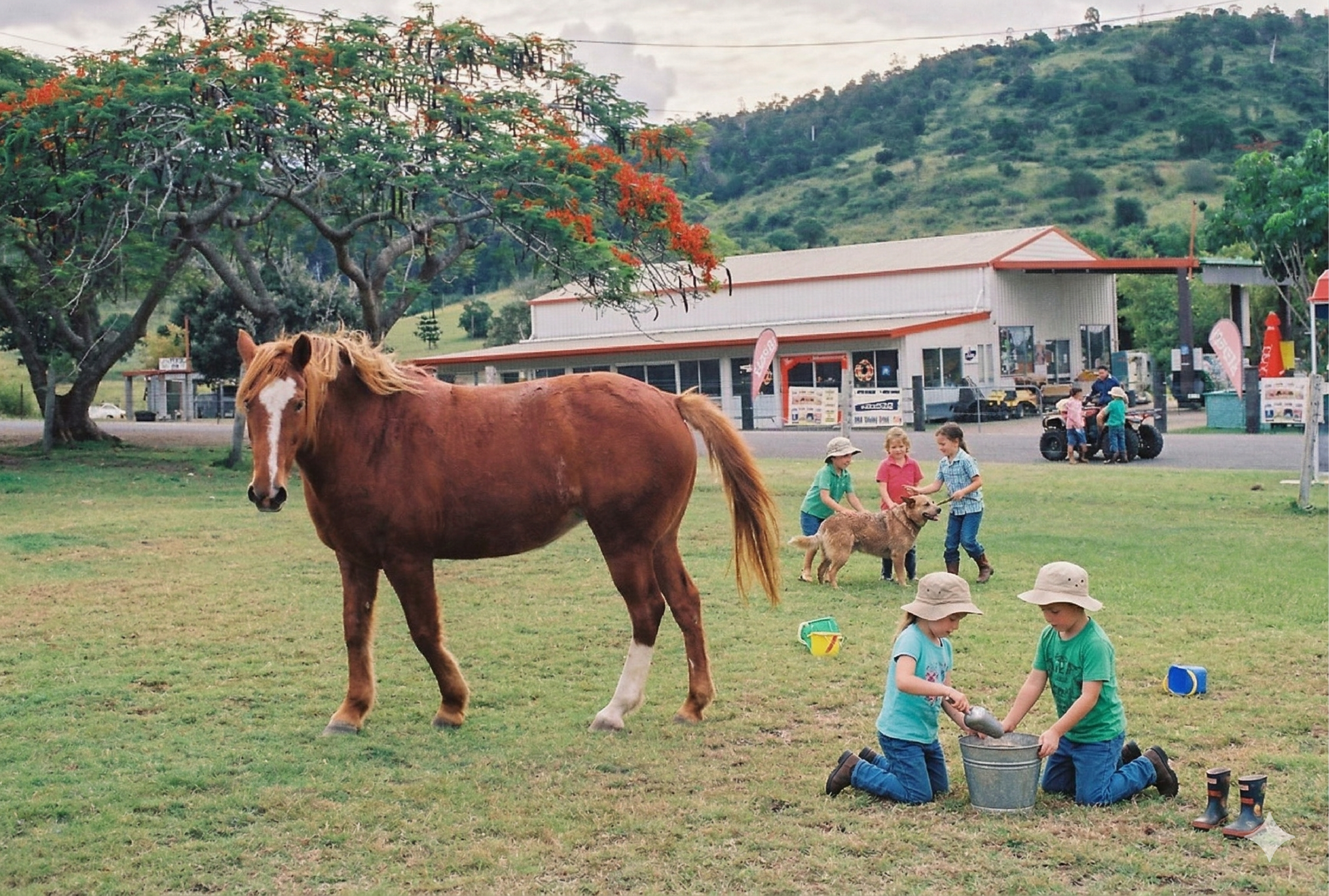}
     \caption{An example image to show that the importance of semantic components is dynamic and dependent on the user's intent.}
     \label{fig:scene}
 \end{figure}

Second, to minimize the semantic distortion due to wireless transmissions, some recent works in semantic coding have proposed resource allocation algorithms which can prioritize semantically important features~\cite{Park2026}. For example, high-importance features can either be mapped to subcarriers with better channel conditions in orthogonal frequency-division multiplexing (OFDM)~\cite{Liu2024}, or be assigned to stronger spatial channels in multiple-input multiple-output (MIMO) systems using singular value decomposition (SVD)-based precoding~\cite{Weng2024}. Although these approaches are effective in small-scale MIMO systems, they may not be effective in large-scale MIMO systems~\cite{Wang2024LMIMO}. This is because in large-scale MIMO systems, fully digital precoding is impractical due to high hardware cost and power consumption. To overcome this limitation, hybrid beamforming (HBF), which combines low-dimensional digital precoding with high-dimensional analog beamforming, is widely adopted~\cite{Mol2017}. Moreover, large-scale MIMO is typically combined with OFDM to combat frequency-selective fading~\cite{Soh2017}. Despite its importance, the design of HBF schemes that can dynamically prioritize semantically important features in large-scale MIMO-OFDM systems remains an open challenge.

Third, traditional semantic coding approaches, such as DeepSC~\cite{Xie2021} and DeepJSCC~\cite{Bourt2019}, are trained end-to-end on specific datasets, which can lead to overfitting and degraded performance when evaluated on unseen data distribution. This limitation hinders the scalability of semantic coding in real-world applications. While modern large artificial intelligence (AI) models (LAMs), such as ChatGPT, Claude, and Gemini, have demonstrated exceptional zero-shot reasoning capabilities in both vision and language processing tasks, leveraging their inherent strengths to overcome the fundamental challenges in traditional semantic coding remains an important open problem. Some studies (e.g.,~\cite{Xu2025,Guo2025,Huang2025MoE,Jiang2025,chen2024}) have leveraged LAMs in semantic coding. For example, the works in~\cite{Guo2025} and~\cite{Jiang2025}  utilized pretrained  BERT~\cite{Devlin2018} and Bootstrapped Language-Image Pretraining (BLIP) models~\cite{Li2022} as semantic encoders to generate features for text and image data, respectively. Some works have further integrated LAMs into semantic coding with user's query. In~\cite{chen2024},  a large language model (LLM) is used to generate answers and select key semantic elements from images based on user's query. However, the proposed framework in~\cite{chen2024} fine-tuned the LLM on the training dataset to improve the performance. Although fine-tuning can enhance the system performance, it can compromise its generalization ability on unseen objects since the model may overfit to the training distribution and lose the broad representations learned during pretraining. In real-world applications, the input data are diverse and may include previously unseen objects, making generalization a critical challenge for semantic coding. In our previous work~\cite{Huang2025MoE}, we explored the zero-shot capability of semantic coding for text transmission, where the semantic content is more explicit and simple. A joint cross-layer design for image semantic coding that can preserve the generalization capability of LAMs and consider the user's intent in large-scale MIMO-OFDM is more promising for future applications but remains largely unexplored.

Based on the aforementioned observations, in this paper, we propose a system-level framework aiming to jointly address the following questions:
\begin{enumerate}[label=Q\arabic*:]
    \item \textit{When transmitting semantically rich content, how to integrate user's intent in semantic coding?}
    \item \textit{How to design a hybrid beamforming scheme that can prioritize semantically important features in large-scale MIMO-OFDM systems?}
    \item \textit{How to improve the generalization capability of traditional semantic coding on unseen object categories?}
\end{enumerate}
The main contributions of this paper are as follows:
\begin{figure*}
    \centering
    \includegraphics[width=\linewidth]{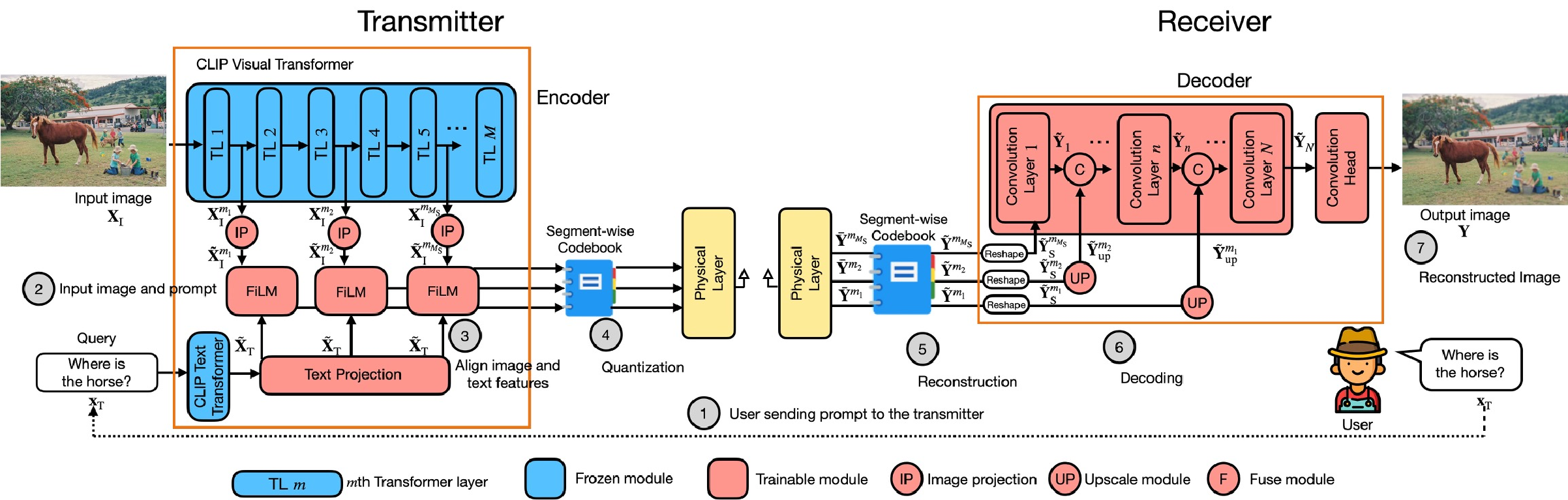}
    \caption{The system model of the proposed generalized QO-ISC framework. The gray numbered circles denote the sequential steps of the communication and processing workflow.}
    \label{fig:system_model}
\end{figure*}
    
\begin{itemize}
   \item We propose a generalized query-oriented image semantic coding (QO-ISC) framework and integrate it with a  semantic-aware hybrid beamforming (SA-HBF) algorithm, which prioritizes the semantic features related to the user's intent in a large-scale MIMO-OFDM system. To incorporate the user's intent, the transmitter receives a text query from the user before transmission. We use Contrastive Language-Image Pre-training (CLIP)~\cite{Alec2021}, which is a pretrained LAM, to improve generalization and extract features representations from both the query and the input image. These features are aligned, partitioned into segments, and quantized by a segment-wise codebook. Moreover, we develop an SA-HBF algorithm that trains a weight module to evaluate the importance of the features. The output weight matrix is then used to design the hybrid beamformer for the transmission of features in large-scale MIMO-OFDM system. The receiver of QO-ISC reconstructs the image based on the received features.
    \item To train the proposed QO-ISC, we develop a three-stage optimization framework. We first introduce a \textit{user-intent relevance loss} that uses a pretrained Large Language-and-Vision Assistant (LLaVA) model~\cite{Liu2023} to evaluate the relevance between the generated image and the original image for a given text query. In Stage I, the total loss includes the user-intent relevance loss, $\ell_1$ loss, quantization loss, and adversarial training~\cite{Good2014}.  This stage addresses question Q1 by enforcing the model to preserve query-related features, and question Q3 by aligning the learned semantic embeddings with the generalizable representations from the frozen CLIP model. Stage II focuses on designing the SA-HBF algorithm that is related to question Q2. We develop a weighting mechanism in SA-HBF and formulate a contextual bandit problem. In each round of the contextual bandit problem, the output weight matrix is used to design the hybrid beamformer via a weighted minimum mean square error (WMMSE) algorithm. This algorithm prioritizes subcarriers which carry important features over wireless channel. 
    In Stage III, the entire communication framework is jointly fine-tuned to further enhance the end-to-end semantic reconstruction performance. 

    \item For performance evaluation, we train our proposed generalized QO-ISC framework on the subset of visual question answering (VQA) dataset~\cite{Stan2015} and perform zero-shot evaluation on the subset only containing unseen objects. We compare the answers generated by BLIP-2~\cite{Li2022} between the reconstructed and original images with the same query. For open-ended questions, we use BERTScore~\cite{Zhang2020} to evaluate the similarity of both answers. For all other question types, we use the exact match. Results show that our framework demonstrates superior robustness in preserving query-related features, even in low SNR scenario. When the SNR is equal to $-25$ dB, our proposed generalized QO-ISC framework outperforms the variant using equal weight in SA-HBF by 4.8\% and the version without query alignment by 9\% in terms of answer match rate. When compared with the state-of-the-art query-based image semantic coding approach~\cite{chen2024}, our proposed framework achieves an 8\% gain. It also demonstrates a 30\% improvement over the state-of-the-art joint source-channel coding (JSCC) codec, SwinJSCC~\cite{Yang2025}.    
    
\end{itemize}
This paper is organized as follows. Section~\ref{Sec:system} presents the system model, including the generalized QO-ISC framework and the SA-HBF design. Section~\ref{Sec:problem} presents the problem formulation and a three-stage optimization framework. Section~\ref{Sec:training} describes the training procedures for the encoder, decoder, and codebook. Section~\ref{Sec:beamforming} describes the design and optimization of the SA-HBF. Section~\ref{Sec:simulation} evaluates the performance of the proposed generalized QO-ISC framework through simulations. Conclusions are given in Section~\ref{Sec:conclude}. The WMMSE hybrid beamforming algorithm is presented in the Appendix.

\textit{Notations:} We use boldface lowercase letters to denote vectors, boldface uppercase letters to denote matrices, and calligraphic letters to represent sets. The transpose, Hermitian transpose, inverse, and vectorization of matrix $\mathbf{X}$ are denoted by $\mathbf{X}^{T}$, $\mathbf{X}^{H}$, $\mathbf{X}^{-1}$, and $\operatorname{vec}(\mathbf{X})$, respectively. The $i$th row of matrix $\mathbf{X}$ is denoted by $\mathbf{x}_i$, and the $(i,j)$th element by $[\mathbf{X}]_{ij}$. $\mathcal{CN}(\boldsymbol{\mu}, \boldsymbol{\Sigma})$ denotes the complex Gaussian distribution, where $\boldsymbol{\mu}$ and $\boldsymbol{\Sigma}$ are the mean vector and covariance matrix, respectively. The $\ell_1$-norm, $\ell_2$-norm, and Frobenius norm are denoted by $\|\cdot\|_1$, $\|\cdot\|_2$, and $\|\cdot\|_{F}$, respectively. The expectation operator is represented by $\mathbb{E}[\cdot]$, $\odot$ denotes element-wise multiplication, and the dot product between vectors $\mathbf{a}$ and $\mathbf{b}$ is denoted by $\langle \mathbf{a}, \mathbf{b} \rangle$. 

\section{System Model}
\label{Sec:system}
Our proposed generalized QO-ISC framework is illustrated in Fig.~\ref{fig:system_model}. It consists of a transmitter and a receiver equipped with $N_{\text{t}}$ and $N_{\text{r}}$ antennas, respectively. The transmitter obtains an image from a visual sensor. Before transmission, the user provides a text query that specifies the intended semantic content (e.g., a horse) to the transmitter based on general knowledge of the scene, such as the layout of the area or the typical presence of certain people or objects. While the user has general knowledge of the environment, the query narrows the focus onto specific semantic elements. Our proposed framework prioritizes those specific elements during transmission. We assume reliable transmission of the text query\footnote{Reliable transmission is feasible because the text query is substantially smaller in size than the image features. It can be transmitted over a dedicated control channel prior to image transmission. By using channel coding (e.g.,~\cite{3gpp_ts36212}), the query can be transmitted in a reliable manner.}. The transmitter then takes both the image and the text query as input. It extracts visual features from the input image and text features from the query. These features are aligned together to obtain semantic features which are relevant to the user's intent. Since different features do not contribute equally to the final result, a weight module is used to assign different importance weight to the features. The features are then quantized using a segment-wise codebook. The SA-HBF algorithm prioritizes the transmission of features which have higher importance weight. At the receiver, the decoder reconstructs the image from the received features. The goal of the generalized QO-ISC framework is to reconstruct the image while prioritizing the semantic content relevant to the user's intent, as indicated by the query. 
\begin{figure*}[t]
    \centering
    \includegraphics[width=\linewidth]{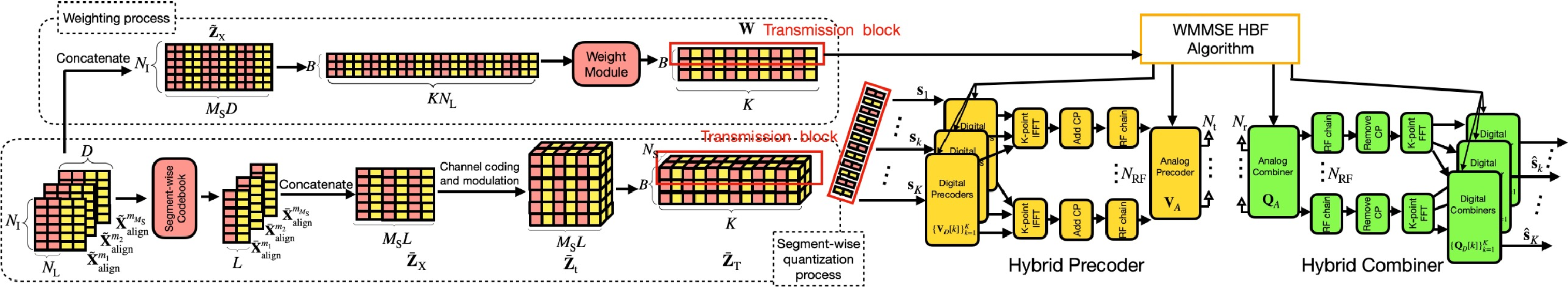}
    \caption{The weighting process, segment-wise quantization process, and the hybrid beamformer in the generalized QO-ISC framework. As an example, we consider $M_{\text{S}}=3$ feature matrices, each with dimensions $N_{\text{I}}=6$ and $D=4$. With a segment length of $N_{\text{L}}=2$, each row of the feature matrix is partitioned into $L=D/N_{\text{L}}=2$ segments. Before transmission, the features are concatenated and reshaped into $B=3$ transmission blocks. Each transmission block is mapped to $K=12$ subcarriers and $N_{\text{S}}=2$ data symbols per subcarrier.}
    \label{fig:transfer}
\end{figure*}

Specifically, in the transmitter, we leverage CLIP~\cite{Alec2021}, which is an LAM pretrained on extensive image-text pairs, to extract the features from the text query and the input image. Let $\mathbf{x}_{\text{T}}$ denote the input text query. Let $\mathbf{X}_{\text{I}} \in \mathbb{R}^{C \times H \times W}$ denote the input image, where $C$, $H$, and $W$ are the number of channels, the dimensions along the height and width directions, respectively. The text query is processed by the CLIP text transformer to extract the text feature matrix $\mathbf{X}_{\text{T}} \in \mathbb{R}^{N_{\text{T}} \times D}$, where $N_{\text{T}}$ is the number of text embeddings and $D$ is the embedding dimension. The image is processed by the CLIP visual transformer, which consists of $M$ transformer layers. For each layer $m \in \mathcal{M} = \{1, 2, \ldots, M\}$, the output feature matrix is denoted by $\mathbf{X}_{\text{I}}^{m} \in \mathbb{R}^{N_{\text{I}} \times D}$, where $N_{\text{I}}$ is the number of image embeddings. Inspired by skip connections in U-Net~\cite{Ronne2015}, we forward the features from different encoder layers to the decoder in order to preserve the spatial details for better reconstruction. However, transmitting the features from all transformer layers can incur a large transmission overhead. Thus, we only select a subset of transformer layers indexed by $\mathcal{M}_{\text{S}}=\{m_1,\dots,m_{M_{\text{S}}}\} \subseteq \mathcal{M}$, where $M_{\text{S}}=|\mathcal{M}_{\text{S}}| \leq M $. Only the feature matrices $\mathbf{X}_{\text{I}}^{m}$, for $m\in \mathcal{M}_{\text{S}}$, are processed in the subsequent transmission pipeline. Note that all transformer layers in the CLIP visual and text transformers are kept frozen to preserve the pretrained model's generalization ability.
To extract the image features relevant to the user's intent, we align the text and image features by projecting both modalities into a shared embedding space using a multilayer perceptron (MLP). 
The projected image and text feature matrices are given by:
\begin{align}
    \mathbf{\tilde{X}}_{\text{I}}^{m} &= \mathcal{P}_{\text{I}}^{m}(\mathbf{X}_{\text{I}}^{m}; \boldsymbol{\theta}^{m}_{\text{P, I}}) \in \mathbb{R}^{N_{\text{I}} \times D}, \quad m\in \mathcal{M}_{\text{S}}, \\
    \mathbf{\tilde{X}}_{\text{T}} &= \mathcal{P}_{\text{T}}(\mathbf{X}_{\text{T}}; \boldsymbol{\theta}_{\text{P, T}}) \in \mathbb{R}^{N_{\text{T}} \times D},
\end{align}
where $\mathcal{P}_{\text{I}}^{m}(\cdot;\, \boldsymbol{\theta}^{m}_{\text{P, I}})$ and $\mathcal{P}_{\text{T}}(\cdot;\, \boldsymbol{\theta}_{\text{P, T}})$ are the projection functions for the $m$th selected image features and text features with parameter sets $\boldsymbol{\theta}^{m}_{\text{P, I}}$ and $\boldsymbol{\theta}_{\text{P, T}}$, respectively. Each projected image feature matrix $\mathbf{\tilde{X}}_{\text{I}}^{m}$ is paired with the projected text feature matrix $\mathbf{\tilde{X}}_{\text{T}}$. They are processed by Feature-wise Linear Modulation (FiLM)~\cite{Dum2018} to align the two modalities. Specifically, $\mathbf{\tilde{X}}_{\text{T}}$ is passed through two MLPs to generate the scale matrix $\mathbf{\tilde{X}}_{\text{T,sc}}^{m}\in \mathbb{R}^{N_{\text{I}} \times D}$ and shift matrix $\mathbf{\tilde{X}}_{\text{T,sh}}^{m} \in \mathbb{R}^{N_{\text{I}} \times D}$: 
\begin{align}
    \mathbf{\tilde{X}}_{\text{T,sc}}^{m} = \mathcal{E}_{\text{sc}}^{m}(\mathbf{\tilde{X}}_{\text{T}};\boldsymbol{\theta}_{\text{sc}}^{m}),~ 
    \mathbf{\tilde{X}}_{\text{T,sh}}^{m} = \mathcal{E}_{\text{sh}}^{m}(\mathbf{\tilde{X}}_{\text{T}};\boldsymbol{\theta}_{\text{sh}}^{m}),
    \end{align}
where $\mathcal{E}_{\text{sc}}^{m}(\cdot;\boldsymbol{\theta}_{\text{sc}}^{m})$ and $\mathcal{E}_{\text{sh}}^{m}(\cdot;\boldsymbol{\theta}_{\text{sh}}^{m})$ denote the MLP networks for scale matrix and shift matrix, respectively, and $\boldsymbol{\theta}_{\text{sc}}^{m}$ and $\boldsymbol{\theta}_{\text{sh}}^{m}$ denote their learnable parameter sets. The aligned feature matrix $\mathbf{\tilde{X}}_{\text{align}}^{m} \in \mathbb{R}^{N_{\text{I}} \times D}$ is given by
\begin{align} 
    \mathbf{\tilde{X}}_{\text{align}}^{m} = \mathbf{\tilde{X}}_{\text{T,sc}}^{m} \odot \mathbf{\tilde{X}}_{\text{I}}^{m}+\mathbf{\tilde{X}}_{\text{T,sh}}^{m}, \quad m\in \mathcal{M}_{\text{S}}. \label{eq:align}
\end{align} 
Note that equation~(\ref{eq:align}) performs an affine transformation to the image features. If the semantic features of the input image are not related to the query, then the emphasis on query-related features will be weakened, and the aligned features will remain close to the original image representation $\mathbf{\tilde{X}}_{\text{I}}^{m},m\in \mathcal{M}_{\text{S}}$.

To further compress the features, we apply vector quantization to convert each embedding vector in matrix $\mathbf{\tilde{X}}_{\text{align}}^{m}$ into a sequence of discrete integers~\cite{Oord2017}. We partition each embedding vector into multiple segments and quantize each segment independently using its own set of codewords. For an embedding vector with dimension $D$, we partition it into $L$ segments, where each segment has length $N_{\text{L}}=\frac{D}{L}$. We assume $D$ is divisible by $L$, so that the embedding vector can be partitioned into segments of equal length. The $l$th codebook $\mathcal{C}_{l} = \left\{ \mathbf{c}_i^l \in \mathbb{R}^{1 \times N_{\text{L}}} \right\}_{i=1}^{N_{\text{cw}}}$ consists of a set of codewords $\mathbf{c}_i^l$, where $N_{\text{cw}}$ is the number of codewords. For the aligned feature matrix $\mathbf{\tilde{X}}_{\text{align}}^{m} \in \mathbb{R}^{N_{\text{I}} \times D}$, we obtain the quantized feature matrix $\mathbf{\bar{X}}_{\text{align}}^{m}$ by using the segment-wise vector quantization function $\mathcal{Q}_{\text{seg}}(\cdot;\mathcal{C})$, where $\mathcal{C} = \{ \mathbf{c}_i^{l} \, |\, i = 1, \ldots, N_{\text{cw}},\ l = 1, \ldots, L \}$ is the set of all codebooks. We have
\begin{align}
    \mathbf{\bar{X}}_{\text{align}}^{m} = \mathcal{Q}_{\text{seg}}(\mathbf{\tilde{X}}_{\text{align}}^{m}; \mathcal{C}) \in \{1, \ldots, N_{\text{cw}}\}^{N_{\text{I}} \times L}, m\in \mathcal{M}_{\text{S}}. \label{eq:quant}
\end{align}
The segment-wise vector quantization partitions each row feature vector $\mathbf{\tilde{x}}_{\text{align}}^{m} \in \mathbb{R}^{1 \times D}$ in the aligned feature  matrix $\mathbf{\tilde{X}}_{\text{align}}^{m}$ into $L$ segments of length $N_{\text{L}}$, and assigns each segment to its nearest codeword in codebook $\mathcal{C}_{l}$. The operation of function  $\mathcal{Q}_{\text{seg}}(\mathbf{\tilde{X}}_{\text{align}}^{m}; \mathcal{C})$ is as follows. For each row index $n \in \{1, \ldots, N_{\text{I}}\}$ and segment index $l \in \{1, \ldots, L\}$, we extract the $l$th segment from the $n$th row vector of $\mathbf{\tilde{X}}_{\text{align}}^{m}$ as
\begin{align}
    \mathbf{u}_{n,l} = \big( 
        &[\mathbf{\tilde{X}}_{\text{align}}^{m}]_{n,\,(l-1)N_{\text{L}} + 1},\ 
         [\mathbf{\tilde{X}}_{\text{align}}^{m}]_{n,\,(l-1)N_{\text{L}} + 2},\ \notag \\
        &\dots,\ 
         [\mathbf{\tilde{X}}_{\text{align}}^{m}]_{n,\,(l-1)N_{\text{L}} + N_{\text{L}}} 
    \big) \in \mathbb{R}^{1 \times N_{\text{L}}}. \label{eq:segment}
\end{align}
We obtain the quantized matrix $\mathbf{\bar{X}}_{\text{align}}^{m}$, where each quantized index is given by
\begin{align}
    \left[ \mathbf{\bar{X}}_{\text{align}}^{m} \right]_{n,l} &= 
    \arg\min_{i \in \{1, \ldots, N_{\text{cw}}\}} 
    \left\| \mathbf{u}_{n,l} - \mathbf{c}_i^l \right\|_2^2, \notag \\
    &\quad n \in \{1, \ldots, N_{\text{I}}\},\;
    l \in \left\{1, \ldots,L \right\}. \label{eq:quant_index}
\end{align}
As illustrated in Fig.~\ref{fig:transfer}, we horizontally concatenate the quantized feature matrices $\{\mathbf{\bar{X}}_{\text{align}}^{m}\}_{m \in \mathcal{M}_{\text{S}}}$ after quantization to construct the matrix $\mathbf{\bar{Z}}_{\text{X}}$, which is given by
\begin{align}
    \mathbf{\bar{Z}}_{\text{X}} = 
    \left[ 
    \mathbf{\bar{X}}_{\text{align}}^{m_1} \ 
    \mathbf{\bar{X}}_{\text{align}}^{m_2} \ 
    \cdots \ 
    \mathbf{\bar{X}}_{\text{align}}^{m_{M_{\text{S}}}}
    \right]
    \in \{1, \ldots, N_{\text{cw}}\}^{N_{\text{I}} \times (M_{\text{S}} L)}.
\end{align}
The matrix $\mathbf{\bar{Z}}_{\text{X}}$ is then passed through channel coding and modulation processes in physical layer. The resulting complex-valued transmit signal is denoted by $\mathbf{\bar{Z}}_{\text{t}} \in \mathbb{C}^{N_{\text{I}} \times (M_{\text{S}} L)\times N_{\text{P}}}$, where $N_{\text{P}}$ represents the number of modulated symbols per feature element after channel coding and modulation. To prepare the signal for transmission over a large-scale MIMO-OFDM system, we reshape the modulated tensor $\mathbf{\bar{Z}}_{\text{t}}$ according to the time-frequency structure of the physical layer. In the OFDM framework, the system operates over $K$ subcarriers, with $N_{\text{S}}$ data streams transmitted per subcarrier. We segment $\mathbf{\bar{Z}}_{\text{t}}$ into $B$ transmission blocks, where each block is shaped to span a full $K\times N_{\text{S}}$ time-frequency grid. Each transmission block is transmitted sequentially using SA-HBF. We reshape the modulated feature tensor $\mathbf{\bar{Z}}_{\text{t}}$ into $\mathbf{\bar{Z}}_{\text{T}} \in \mathbb{C}^{B \times K \times N_{\text{S}}}$, and transmit each block $\mathbf{\bar{Z}}_{\text{T}, b} \in \mathbb{C}^{K \times N_{\text{S}}}$ in sequence.

Since each feature has different contributions to the final reconstruction, we employ a weight module to evaluate the importance weight of features for each subcarrier within a transmission block. The weight module takes the concatenation of feature matrices before quantization as input, which is defined as 
\begin{align}
    \mathbf{\tilde{Z}}_{\text{X}}=\left[ 
        \mathbf{\tilde{X}}_{\text{align}}^{m_1} \ 
        \mathbf{\tilde{X}}_{\text{align}}^{m_2} \ 
        \cdots \ 
        \mathbf{\tilde{X}}_{\text{align}}^{m_{M_{\text{S}}}}
        \right]
        \in \mathbb{R}^{N_{\text{I}} \times (M_{\text{S}} D)}.
\end{align}
To evaluate the relative importance of features across subcarriers, we reshape the feature matrix $\mathbf{\tilde{Z}}_{\text{X}}$ into $\mathbf{\tilde{Z}}_{\text{T}} \in \mathbb{R}^{B \times (K N_{\text{L}})}$. For the features in transmission block $b$, the weight module takes the row vector $\mathbf{\tilde{z}}_{\text{T}, b} \in \mathbb{R}^{1\times (K N_{\text{L}})}$ of $\mathbf{\tilde{Z}}_{\text{T}}$ as input and outputs a weight vector
\begin{align}
    \mathbf{w}_{b} = \mathcal{W}(\mathbf{\tilde{z}}_{\text{T}, b}; \boldsymbol{\psi}_{\text{W}}) \in \mathbb{R}^{1\times K}, \quad b = 1,\dots,B, \label{eq:weight}
    \end{align}
where $\mathcal{W}(\cdot;\, \boldsymbol{\psi}_{\text{W}})$ denotes the function of the weight module with parameter set $\boldsymbol{\psi}_{\text{W}}$. The weight vector $\mathbf{w}_{b} \in {\mathbb{R}}^{1\times K}$ assigns relative importance to each of the $K$ subcarriers in transmission block $b$. Each element $w_{b,k}\in [0,1]$ in weight vector $\mathbf{w}_{b}$ satisfies the normalization condition $\sum_{k=1}^{K} w_{b,k} = 1$. After collecting all the weights for each transmission block, the resulting weight matrix $\mathbf{W} \in  \mathbb{R}^{B \times K}$ is used in the subsequent design of SA-HBF algorithm to prioritize subcarriers carrying important features during transmission.

For the features in transmission block $\mathbf{\bar{Z}}_{\text{T}, b}\in \mathbb{C}^{K \times N_{\text{S}}}$, the data symbols corresponding to subcarrier $k \in \mathcal{K} = \{1, \ldots, K\}$ are denoted by $\mathbf{s}_k \in \mathbb{C}^{N_{\text{S}}}$. As illustrated in Fig.~\ref{fig:transfer}, these data symbols are first processed by a low-dimensional digital precoder $\mathbf{V}_{\text{D}}[k] \in \mathbb{C}^{N_{\text{RF}} \times N_{\text{S}}}$, where $N_{\text{RF}}$ is the number of radio frequency (RF) chains. The precoded signals are then passed through $N_{\text{RF}}$ $K$-point inverse fast Fourier transforms (IFFTs) to convert them from the frequency domain to the time domain. Since practical MIMO-OFDM systems have a limited number of RF chains, we assume $N_{\text{RF}} \ll \min(N_{\text{t}}, N_{\text{r}})$. To mitigate inter-symbol interference (ISI), a cyclic prefix (CP) is inserted to the time-domain signals after the IFFTs. Finally, an analog precoder $\mathbf{V}_{\text{A}} \in \mathbb{C}^{N_{\text{t}} \times N_{\text{RF}}}$ maps the signals onto the large-scale MIMO antenna array for transmission. With hybrid digital and analog precoding, the transmitted signal at subcarrier $k$, which is denoted by $\mathbf{z}_k \in \mathbb{C}^{N_{\text{t}}}$, can be expressed as
\begin{align} 
    \mathbf{z}_{k} = \mathbf{V}[k] \mathbf{s}_{k}, \quad k \in \mathcal{K},
\end{align}
where $\mathbf{V}[k] \!\!=\!\! \mathbf{V}_{\text{A}} \mathbf{V}_{\text{D}}[k]$ is the hybrid precoder at subcarrier $k$.

The transmitted signals propagate through a wireless channel. The received signal $\hat{\mathbf{z}}_{k}\in \mathbb{C}^{N_{\text{r}}}$ at subcarrier $k$ is given by
\begin{align} 
    \hat{\mathbf{z}}_{k} = \mathbf{H}[k] \mathbf{z}_{k} + \mathbf{n}_{k}, \quad k \in \mathcal{K},
\end{align}
where $\mathbf{H}[k] \in \mathbb{C}^{N_{\text{r}} \times N_{\text{t}}}$ is the channel matrix and $\mathbf{n}_k \sim \mathcal{CN}(\mathbf{0}, \sigma^2 \mathbf{I}_{N_{\text{r}}})$ is the noise with zero mean and covariance $\sigma^2 \mathbf{I}_{N_{\text{r}}}$. Specifically, the channel is modeled as a geometry-based multipath fading channel with $N_{\text{C}}$ clusters and $N_{\text{R}}$ rays per cluster. The channel response at subcarrier $k$ is given by
\begin{align} 
    \mathbf{H}[k] = \sqrt{\frac{N_{\text{t}} N_{\text{r}}}{N_{\text{C}} N_{\text{R}}}} \sum_{c=1}^{N_{\text{C}}} \sum_{r=1}^{N_{\text{R}}} \alpha_{c,r} \, \mathbf{a}_r(\theta_{c,r}^{\text{r}}) \, \mathbf{a}_t(\theta_{c,r}^{\text{t}})^{H},
\end{align}
where $\alpha_{c,r} \sim \mathcal{CN}(0,1)$ is the complex path gain of the $r$th ray in the $c$th cluster. The normalized antenna array response vectors at the receiver and transmitter are respectively given by
\begin{align}
    \mathbf{a}_r(\theta_{c,r}^{\text{r}}) &= \frac{1}{\sqrt{N_{\text{r}}}} 
    \begin{bmatrix}
        1 & e^{j\pi \sin \theta_{c,r}^{\text{r}}} & \cdots & e^{j\pi (N_{\text{r}} - 1) \sin \theta_{c,r}^{\text{r}}}
    \end{bmatrix}^{T}, \\
    \mathbf{a}_t(\theta_{c,r}^{\text{t}}) &= \frac{1}{\sqrt{N_{\text{t}}}} 
    \begin{bmatrix}
        1 & e^{j\pi \sin \theta_{c,r}^{\text{t}}} & \cdots & e^{j\pi (N_{\text{t}} - 1) \sin \theta_{c,r}^{\text{t}}}
    \end{bmatrix}^{T},
\end{align}
where $\theta_{c,r}^{\text{r}}$ and $\theta_{c,r}^{\text{t}}$ denote the angle of arrival (AoA) and angle of departure (AoD), respectively, for the $r$th ray in the $c$th cluster.

At the receiver, signals from all subcarriers are first processed by an analog combiner $\mathbf{Q}_{\text{A}} \in \mathbb{C}^{N_{\text{r}} \times N_{\text{RF}}}$ and then passed through $N_{\text{RF}}$ RF chains. After removing the cyclic prefix, $N_{\text{RF}}$ $K$-point fast Fourier transforms (FFTs) are applied to convert the signals back to the frequency domain. Then, a low-dimensional digital combiner $\mathbf{Q}_{\text{D}}[k] \in \mathbb{C}^{N_{\text{RF}} \times N_{\text{S}}}$ is applied to each subcarrier $k$. The processed signal at subcarrier $k$ is given by
\begin{align}
    \hat{\mathbf{s}}_{k} = \mathbf{Q}[k]^H \mathbf{H}[k] \mathbf{z}_{k} + \mathbf{Q}[k]^H \mathbf{n}_{k}, \label{eq:received_signal}
\end{align}
where $\mathbf{Q}[k] = \mathbf{Q}_{\text{A}} \mathbf{Q}_{\text{D}}[k]$ represents the hybrid combiner at subcarrier $k$. The received data streams $\hat{\mathbf{s}}_{k}$ across all subcarriers are aggregated into the feature matrix $\mathbf{\bar{Z}}_{\text{R},b} \in \mathbb{C}^{K \times N_{\text{S}}}$ for the $b$th transmission block. After processing all $B$ transmission blocks, we stack the reconstructed feature matrices to form $\mathbf{\bar{Z}}_{\text{R}}\in \mathbb{C}^{B \times K \times N_{\text{S}}}$. It is then reshaped to $\mathbf{\bar{Z}}_{\text{r}}\in \mathbb{C}^{N_{\text{I}} \times (M_{\text{S}} L)\times N_{\text{P}}}$. After demodulation and channel decoding, we obtain the feature matrix $\mathbf{\bar{Z}}_{\text{Y}} \in \{1, \ldots, N_{\text{cw}}\}^{N_{\text{I}} \times (M_{\text{S}}L)}$. We partition the feature matrix $\mathbf{\bar{Z}}_{\text{Y}}$ along the horizontal axis to obtain the recovered embeddings
\begin{align}
    \mathbf{\bar{Z}}_{\text{Y}} = 
    \left[ 
    \mathbf{\bar{Y}}_{\text{align}}^{m_1} \ 
    \mathbf{\bar{Y}}_{\text{align}}^{m_2} \ 
    \cdots \ 
    \mathbf{\bar{Y}}_{\text{align}}^{m_{M_{\text{S}}}}
    \right]
    \in \{1, \ldots, N_{\text{cw}}\}^{N_{\text{I}} \times (M_{\text{S}} L)},
\end{align}
where each matrix $\mathbf{\bar{Y}}_{\text{align}}^{m} \in \{1, \ldots, N_{\text{cw}}\}^{N_{\text{I}} \times L}$.
Then, the segment-wise reconstruction function $\mathcal{Q}_{\text{seg}}^{-1}(\cdot;\, \mathcal{C})$ is applied to each quantized feature matrix $\mathbf{\bar{Y}}_{\text{align}}^{m}$. This function reverses the quantization process by mapping each quantized element into its corresponding codeword in the codebook. Let the element in the $n$th row and $l$th column of matrix $\mathbf{\bar{Y}}_{\text{align}}^{m}$ be denoted by $\bar{y}_{n,l}^{m} \in \{1, \ldots, N_{\text{cw}}\}$. The reconstructed embedding matrix, denoted by $\mathbf{\tilde{Y}}_{\text{align}}^{m} = \mathcal{Q}^{-1}(\bar{\mathbf{Y}}_{\text{align}}^{m};\, \mathcal{C})$, is given by
\begin{align}
    \mathbf{\tilde{Y}}_{\text{align}}^{m} =
    \begin{bmatrix}
        \mathbf{c}^{1}_{\bar{y}_{1,1}^{m}} & \mathbf{c}^{2}_{\bar{y}_{1,2}^{m}} & \cdots & \mathbf{c}^{L}_{\bar{y}_{1,L}^{m}} \\
        \mathbf{c}^{1}_{\bar{y}_{2,1}^{m}} & \mathbf{c}^{2}_{\bar{y}_{2,2}^{m}} & \cdots & \mathbf{c}^{L}_{\bar{y}_{2,L}^{m}} \\
        \vdots & \vdots & \ddots & \vdots \\
        \mathbf{c}^{1}_{\bar{y}_{N_{\text{I}},1}^{m}} & \mathbf{c}^{2}_{\bar{y}_{N_{\text{I}},2}^{m}} & \cdots & \mathbf{c}^{L}_{\bar{y}_{N_{\text{I}},L}^{m}}
    \end{bmatrix}
    \in \mathbb{R}^{N_{\text{I}} \times D},
    \label{eq:dequant_matrix}
\end{align}
where $\mathbf{c}^{l}_{\bar{y}_{n,l}^{m}} \in \mathbb{R}^{1 \times N_{\text{L}}}$ is the codeword associated with index $\bar{y}_{n,l}^{m}$ from the $l$th codebook $\mathcal{C}_{l}= \{ \mathbf{c}_i^l \in \mathbb{R}^{1 \times N_{\text{L}}} \}_{i = 1}^{N_{\text{cw}}}$. 

After obtaining the reconstructed embeddings $\mathbf{\tilde{Y}}_{\text{align}}^{m}$ for all $m \in \mathcal{M}_{\text{S}}$, each matrix is reshaped into a spatial feature tensor $\mathbf{\tilde{Y}}_{\text{S}}^{m} \in \mathbb{R}^{\tilde{H} \times \tilde{W} \times D}$, where $\tilde{H}\tilde{W}=N_{\text{I}}$. These spatial feature matrices are then passed through a decoder composed of $N$ convolution layers, indexed by $n \in \mathcal{N}=\{1, \ldots, N\}$. Each convolutional layer is represented by function $\mathcal{D}_{n}(\cdot;\, \boldsymbol{\phi}_{\text{D}}^{n})$, parameterized by $\boldsymbol{\phi}_{\text{D}}^{n}$, and produces an output tensor $\mathbf{\tilde{Y}}^{n}\in \mathbb{R}^{\tilde{H}_{n} \times \tilde{W}_{n} \times D}$, where the spatial dimensions $\tilde{H}_{n}$ and $\tilde{W}_{n}$ are progressively increased in each layer. Each convolutional block consists of two standard convolution layers and a transposed convolution to perform spatial upsampling, where each standard convolution layer is followed by instance normalization and a leaky rectified linear unit (ReLU) activation. The number of output channels is equal to $D$ for all layers. The decoder processes the embeddings in reverse order, starting from the reconstructed feature matrix of the final transformer layer's output $\mathbf{\tilde{Y}}_{\text{S}}^{m_{M_{\text{S}}}}$. For $n = 1$, the feature tensor $\mathbf{\tilde{Y}}_{\text{S}}^{m_{M_{\text{S}}}}$ is passed directly to the first convolutional layer. \begin{samepage}For $n>1$, the output from the previous layer $\mathbf{\tilde{Y}}^{n-1}$ is fused with the upsampled version of the corresponding reconstructed embedding, defined as $ \mathbf{\tilde{Y}}_{\text{up}}^{m_{M_{\text{S}} - n + 1}} = \mathcal{U}_{n}( \mathbf{\tilde{Y}}_{\text{S}}^{m_{M_{\text{S}} - n + 1}}; \boldsymbol{\phi}_{\text{U}}^{n} )$, where $\mathcal{U}_{n}(\cdot; \boldsymbol{\phi}_{\text{U}}^{n})$ denotes a learnable upsampling module.\end{samepage} This upsampled tensor is then fused with the output $\mathbf{\tilde{Y}}^{n-1}$ via a feature fusion MLP, defined as $\mathbf{\tilde{Y}}_{\text{fused}}^{n-1} = \mathcal{F}_{n}\left( \mathbf{\tilde{Y}}_{\text{up}}^{m_{M_{\text{S}} - n + 1}}, \mathbf{\tilde{Y}}^{n-1};\boldsymbol{\phi}_{\text{F}}^{n} \right)$ where $\mathcal{F}_{n}(\cdot, \cdot; \boldsymbol{\phi}_{\text{F}}^{n})$ merges the aligned features and intermediate output. The fused tensor $\mathbf{\tilde{Y}}_{\text{fused}}^{n-1}$ is then passed to the $n$th convolutional layer
\begin{align}
    \mathbf{\tilde{Y}}^{n} = 
    \begin{cases}
        \mathcal{D}_{n}\left( \mathbf{\tilde{Y}}_{\text{S}}^{m_{M_{\text{S}}}};\, \boldsymbol{\phi}_{\text{D}}^{n} \right), & n = 1, \\[1ex]
        \mathcal{D}_{n}\left( \mathbf{\tilde{Y}}_{\text{fused}}^{n-1};\, \boldsymbol{\phi}_{\text{D}}^{n} \right), & n \in \mathcal{N} \setminus \{1\}.
    \end{cases}
    \label{eq:decoder}
\end{align}
In our work, the number of convolution layers $N$ is equal to the number of selected indices $M_{\text{S}}$ so that each transmitted feature can combine with the convolution output. The final decoded embeddings $\mathbf{\tilde{Y}}^{N}$ pass through the final convolution head, generating the reconstructed image $\mathbf{Y}$. Note that all the trainable parameters in the encoder are included in the parameter set $\boldsymbol{\theta} = \{ \boldsymbol{\theta}^{m}_{\text{P, I}}, \boldsymbol{\theta}_{\text{P, T}}, \boldsymbol{\theta}^{m}_{\text{E,sc}}, \boldsymbol{\theta}^{m}_{\text{E,sh}} \}_{m\in\mathcal{M}_{\text{S}}}$ and all the trainable parameters in the decoder are included in the parameter set $\boldsymbol{\phi} = \{ \boldsymbol{\phi}^{n}_{\text{U}}, \boldsymbol{\phi}^{n}_{\text{F}}, \boldsymbol{\phi}^{n}_{\text{D}}\}_{n\in\mathcal{N}}$.


\section{Problem Formulation} 
\label{Sec:problem} 
 \begin{figure}
     \centering
     \includegraphics[width=\linewidth]{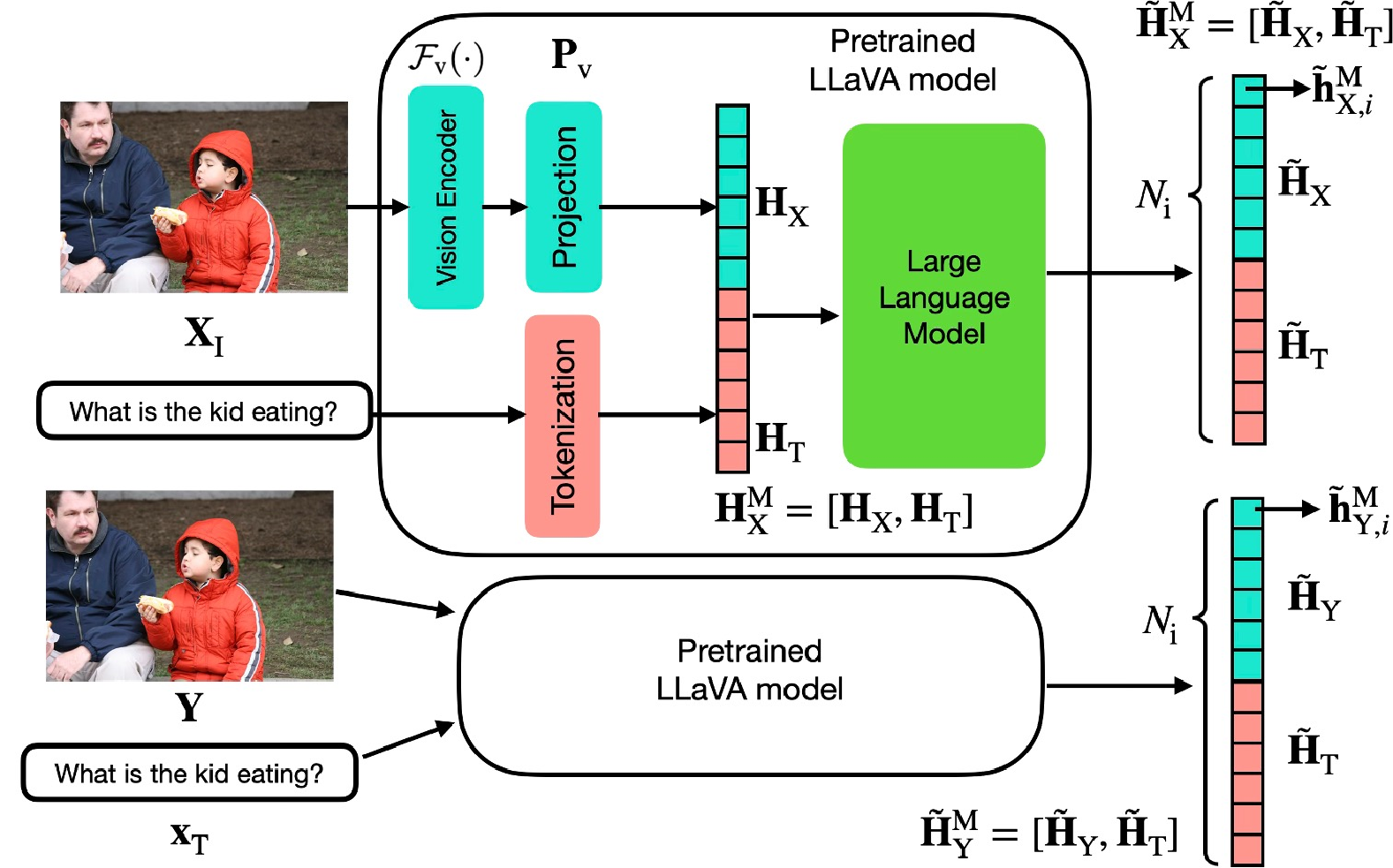}
     \caption{The user-intent relevance loss derived from pretrained LLaVA.}
     \label{fig:LLaVA}
 \end{figure}
The proposed generalized QO-ISC framework reconstructs the source image while prioritizing the semantic features relevant to the user's intent, where the intent is specified by a text query. To evaluate whether the reconstructed image preserves the intended semantics, we compare it with the original image under the same query. We define the \textit{user-intent relevance loss $ \mathcal{L}_{\text{user}}$}, which computes the cosine similarity between the feature representations of the reconstructed and original images extracted by the pretrained LAM. To assess semantic alignment, both the original image $\mathbf{X}_{\text{I}}$ and the reconstructed image $\mathbf{Y}$ are paired with the same text query $\mathbf{x}_{\text{T}}$ and processed separately through a pretrained LAM. The images are processed through the visual encoder $\mathcal{F}_{\text{v}}(\cdot)$ and projection matrix $\mathbf{P}_{\text{v}}$ to obtain $\mathbf{H}_{\text{X}} = \mathbf{P}_{\text{v}} \mathcal{F}_{\text{v}}(\mathbf{X}_{\text{I}})$ and $\mathbf{H}_{\text{Y}} = \mathbf{P}_{\text{v}} \mathcal{F}_{\text{v}}(\mathbf{Y})$.
The text query is tokenized into $\mathbf{H}_{\text{T}}$, and is concatenated with the image embeddings to form the multimodal sequences $\mathbf{H}_{\text{X}}^{\text{M}}=[\mathbf{H}_{\text{X}}, \mathbf{H}_{\text{T}}]$ and $\mathbf{H}_{\text{Y}}^{\text{M}}=[\mathbf{H}_{\text{Y}}, \mathbf{H}_{\text{T}}]$, which are input to the language model. The output hidden states are denoted by $\mathbf{\tilde{H}}_{\text{X}}^{\text{M}} = [\mathbf{\tilde{H}}_{\text{X}}, \mathbf{\tilde{H}}_{\text{T}}] $ and $\mathbf{\tilde{H}}_{\text{Y}}^{\text{M}} = [\mathbf{\tilde{H}}_{\text{Y}}, \mathbf{\tilde{H}}_{\text{T}}] $. The user-intent relevance loss is defined as the cosine distance between the corresponding image tokens:
\begin{align}
    \mathcal{L}_{\text{user}}(\mathbf{X}_{\text{I}}, \mathbf{Y};\, \mathbf{x}_{\text{T}}) 
    = 1 - \frac{1}{N_{\text{i}}} \sum_{i=1}^{N_{\text{i}}} \frac{
        \langle\mathbf{\tilde{h}}_{\text{X}, i}^{\text{M}}, \mathbf{\tilde{h}}_{\text{Y}, i}^{\text{M}}\rangle
    }{
        \| \mathbf{\tilde{h}}_{\text{X}, i}^{\text{M}} \|_2 
        \| \mathbf{\tilde{h}}_{\text{Y}, i}^{\text{M}} \|_2
    },
    \label{eq:loss_user}
\end{align}
where $\mathbf{\tilde{h}}_{\text{X}, i}^{\text{M}}$ and $\mathbf{\tilde{h}}_{\text{Y}, i}^{\text{M}}$ denote the $i$th token embedding vector from $\mathbf{\tilde{H}}_{\text{X}}^{\text{M}}$ and $\mathbf{\tilde{H}}_{\text{Y}}^{\text{M}}$, respectively, and $N_{\text{i}}$ denotes the total number of token embedding vectors. The lower value of $\mathcal{L}_{\text{user}}$ means that the reconstructed image produces a similar semantic representation as the original image when paired with the same prompt, indicating that it preserves the alignment with the user's query. 
Note that we do not compare the answer from the reconstructed image to the ground-truth answer since this would conflate the reconstruction errors made by the pretrained LAM, making it difficult to clearly evaluate how well the semantics are preserved.To distinguish the user-intent relevance loss $\mathcal{L}_{\text{user}}$ derived from different LAMs, we append the model name to the term and add the model name as a superscript to the notation. For example, the \textit{user-intent relevance loss (LLaVA)}, denoted as $ \mathcal{L}_{\text{user}}^{\text{LLaVA}}$, indicates that LLaVA~\cite{Liu2023} is used as the pretrained LAM to generate the multimodal embeddings $\mathbf{\tilde{H}}_{\text{X}}^{\text{M}}$ and $\mathbf{\tilde{H}}_{\text{Y}}^{\text{M}}$. In training, we employ LLaVA to compute the user-intent relevance loss.

In addition to user-intent relevance loss, the visual quality of the reconstructed image is also important for image semantic coding. To improve the visual quality, we incorporate the $\ell_1$ loss and adversarial loss~\cite{Good2014} into the training objective of QO-ISC. The $\ell_1$ loss, defined as $\mathcal{L}_{\text{1}}(\mathbf{X}_{\text{I}}, \mathbf{Y})= \|\mathbf{X}_{\text{I}} - \mathbf{Y}\|_{1}$, improves the pixel-level fidelity between the source and reconstructed images, and preserves low-level visual details such as edges and structural content. The adversarial loss improves the perceptual realism of the reconstructed images. In adversarial training, a lightweight discriminator $\mathcal{D}_{\text{dis}}(\cdot;\boldsymbol{\psi}_{\text{dis}})$ with parameter set $\boldsymbol{\psi}_{\text{dis}}$ is trained to distinguish real images from the generated ones. The generator, which is our model, is trained to produce outputs that are indistinguishable from real images. The adversarial loss function for the generator is defined as $\mathcal{L}_{\text{gen}}(\mathbf{Y})=-\mathbb{E}[\log\mathcal{D}_{\text{dis}}(\mathbf{Y};\boldsymbol{\psi}_{\text{dis}})] $, which optimizes the generator toward producing images that are indistinguishable from real samples. Additionally, the segment-wise codebook $\mathcal{C}$ introduces a quantization loss. Let $\tilde{\mathbf{c}}^{l} = \text{arg min}_{\mathbf{c} \in \mathcal{C}_l} \|\mathbf{u}_{n,l}^{m} - \mathbf{c}\|_2^2$ denote the nearest codeword in the $l$th segment. Let $\mathbf{u}_{n,l}^{m}$ denote the $l$th segment of the $n$th row vector in matrix $\mathbf{\tilde{X}}_{\text{align}}^{m}$. The quantization loss is defined as
\begin{align}
    \mathcal{L}_{\text{quant}} 
    &= \sum_{m \in \mathcal{M}_{\text{S}}} \sum_{n=1}^{N_{\text{I}}} \sum_{l=1}^{L} \Big( 
    \left\| \text{sg}[\mathbf{u}_{n,l}^{m}] - \tilde{\mathbf{c}}^{l} \right\|_2^2 \notag \\
    &\hspace{5em}
    + \gamma
    \left\| \mathbf{u}_{n,l}^{m} - \text{sg}[\tilde{\mathbf{c}}^{l}] \right\|_2^2 
    \Big),
    \label{eq:L_quant}
\end{align}
where  $\text{sg}[\cdot]$ denotes the stop-gradient operator and $\gamma$ is a tunable parameter. Since the codeword selection function in~\eqref{eq:quant} is non-differentiable, the stop-gradient operator ensures separate updates to the encoder and codebook during backpropagation.

We aim to train the proposed QO-ISC framework by minimizing the total loss $\mathcal{L}_{\text{total}}$, which is defined as
\begin{align}
    \mathcal{L}_{\text{total}}(\mathbf{X}_{\text{I}},\mathbf{Y};\, \mathbf{x}_{\text{T}}) 
    &= \mathcal{L}_{\text{1}}(\mathbf{X}_{\text{I}}, \mathbf{Y}) 
    + \lambda_{\text{user}} \mathcal{L}_{\text{user}}^{\text{LLaVA}}(\mathbf{X}_{\text{I}},\mathbf{Y};\, \mathbf{x}_{\text{T}}) \nonumber \\
    &\quad + \lambda_{\text{gen}} \mathcal{L}_{\text{gen}}(\mathbf{Y})
    + \lambda_{\text{quant}} \mathcal{L}_{\text{quant}},
    \label{eq:L_total}
\end{align}
where $\lambda_{\text{user}} $, $\lambda_{\text{gen}}$, and $\lambda_{\text{quant}}$ are positive tunable parameters. Based on $\mathcal{L}_{\text{total}}$, we jointly optimize the trainable parameters in the encoder $\boldsymbol{\theta}$, decoder $\boldsymbol{\phi}$, weight module $\boldsymbol{\psi}_{\text{W}}$, codebook $\mathcal{C}$, hybrid precoding and combining matrices $\mathbf{V}_{\text{A}}$ and $\mathbf{Q}_{\text{A}}$, and tensors $\mathbf{V}_{\text{D}}$ and $\mathbf{Q}_{\text{D}}$. The optimization problem is formulated as follows:
\begin{subequations}\label{eq:P0}
    \begin{align}
        \underset{
            \substack{
                \boldsymbol{\theta},\, \boldsymbol{\phi},\, \boldsymbol{\psi}_{\text{W}},\,\mathcal{C},\, \\
                \mathbf{V}_{\text{A}},\, \mathbf{V}_{\text{D}},\,\mathbf{Q}_{\text{A}},\, \mathbf{Q}_{\text{D}}
            }
        }{\text{minimize}} \quad &
        \mathcal{L}_{\text{total}}(\mathbf{X}_{\text{I}},\mathbf{Y};\, \mathbf{x}_{\text{T}}) \label{eq:P0_objective} \\
        \text{subject to} \quad &
        \|\mathbf{V}[k]\|_{F}^{2} \leq P, \quad k = 1, \ldots, K, \label{eq:P0_constraint1} \\
        & \left|[\mathbf{V}_{\text{A}}]_{ij}\right|^2  = 1, \quad 
        i = 1, \ldots, N_{\text{t}}, \notag \\
        & \hspace{7em}  j = 1, \ldots, N_{\text{RF}}, \label{eq:P0_constraint2} \\
        & \left|[\mathbf{Q}_{\text{A}}]_{ml}\right|^2 = 1, \quad 
        m = 1, \ldots, N_{\text{r}}, \notag \\
        & \hspace{7em} l = 1, \ldots, N_{\text{RF}}. \label{eq:P0_constraint3}
    \end{align}
\end{subequations}
Constraint~\eqref{eq:P0_constraint1} enforces the transmit power limit $P$. Since the analog precoder $\mathbf{V}_{\text{A}}$ and analog combiner $\mathbf{Q}_{\text{A}}$ are implemented using phase shifters, which can only adjust the phases of the input signals, they are subject to the constant modulus constraints in~\eqref{eq:P0_constraint2} and~\eqref{eq:P0_constraint3}. Problem~\eqref{eq:P0} is a nonconvex optimization problem with coupled deep learning parameter sets $\{\boldsymbol{\theta}, \boldsymbol{\phi}, \mathcal{C}\}$ and constant modulus constraints~\eqref{eq:P0_constraint2} and~\eqref{eq:P0_constraint3}, making it challenging to directly obtain an optimal solution. Similar to other recent works (e.g.,~\cite{Huang2023, Zhang2025}), we develop a multi-stage approach to obtain the solution.

\textbf{Stage I: Encoder, Decoder, and Codebook Training.} 
In the first stage, the encoder, decoder, and codebook are optimized to minimize the total loss $\mathcal{L}_{\text{total}}$. To isolate the training of the encoder, decoder, and codebook, the weight module, hybrid precoding and combining matrices are deactivated. The data streams are transmitted independently with equal weight and without spatial processing. This setting allows the semantic encoder and decoder to learn quantized embeddings before integrating the beamforming components.

\textbf{Stage II: Semantic-Aware Hybrid Beamforming Design.}
This stage focuses on optimizing the weight module and hybrid precoding and combining matrices in SA-HBF algorithm. During this stage, the parameters of the encoder $\boldsymbol{\theta}$, decoder $\boldsymbol{\phi}$, and codebook $\mathcal{C}$ trained in Stage I are kept frozen. We formulate a contextual bandit problem~\cite{Latt2020}, where an agent assigns importance weight to each subcarrier based on the features to be transmitted using that subcarrier. For each weight matrix produced by the weight module, we propose an WMMSE HBF algorithm that optimizes the hybrid beamforming matrices to prioritize subcarriers with higher weight. This stage reduces the distortion in semantic embeddings learned in Stage~I. 


\textbf{Stage III: Joint Fine-Tuning of All Modules.} This stage involves jointly optimizing all the trainable parameters and hybrid beamforming matrices to minimize the total loss $\mathcal{L}_{\text{total}}$.

\begin{algorithm}[t]
   \small
   \caption{Three-Stage Training Algorithm for Proposed Generalized QO-ISC}
   \label{alg:three-stage training}
   \begin{algorithmic}[1]
   \State \textbf{Input:} Training dataset of image–text pairs $(\mathbf{X}_{\text{I}}, \mathbf{x}_{\text{T}})$; number of epochs $N_1$, $N_2$, $N_3$ for Stages I–III; learning rates $\eta_{\text{gen}}$, $\eta_{\text{dis}}$; regularization weights $\lambda_{\text{user}}, \lambda_{\text{gen}}, \lambda_{\text{quant}}$; scaling factor $\gamma$.
   \State \textbf{Stage I: Encoder, Decoder, and Codebook Training}
   \State Freeze the parameters of pretrained CLIP in the encoder.
   \State \text{Deactivate the weight module and hybrid beamforming.}
   \State Use Algorithm~\ref{alg:stage1} to train the parameters $\boldsymbol{\theta}$, $\boldsymbol{\phi}$, $\mathcal{C}$, the encoder, decoder, and codebook, respectively.
   \State \textbf{Stage II: Semantic-Aware Hybrid Beamforming Design}
   \State Load $\boldsymbol{\theta}$, $\boldsymbol{\phi}$, $\mathcal{C}$ trained in Stage I and freeze.
   \State Activate the weight module and hybrid beamforming. 
   \State Use Algorithm~\ref{alg:stage2} to train the SA-HBF.
   \State \textbf{Stage III: Joint Fine-Tuning of All Modules}
   \State Load $\boldsymbol{\theta}$, $\boldsymbol{\phi}$, $\mathcal{C}$, $\boldsymbol{\psi}_{\text{W}}$, $\mathbf{V}_{\text{A}}$, $\mathbf{V}_{\text{D}}$, $\mathbf{Q}_{\text{A}}$, $\mathbf{Q}_{\text{D}}$ trained in Stages I and II and unfreeze.
   \For{$i \gets 1$ \textbf{to} $N_3$} 
       \State Sample image–text pairs $(\mathbf{X}_{\text{I}}, \mathbf{x}_{\text{T}})$ from training dataset.
       \State Compute the total loss $\mathcal{L}_{\text{total}}(\mathbf{X}_{\text{I}}, \mathbf{Y};\, \mathbf{x}_{\text{T}})$ from~(\ref{eq:L_total}).
       \State Update parameters $\boldsymbol{\theta}$, $\boldsymbol{\phi}$, $\mathcal{C}$, $\boldsymbol{\psi}_{\text{W}}$, $\mathbf{V}_{\text{A}}$, $\mathbf{V}_{\text{D}}$, $\mathbf{Q}_{\text{A}}$, $\mathbf{Q}_{\text{D}}$ by gradient descent.
   \EndFor
   \State \textbf{Output:} Optimized model parameters: $\{\boldsymbol{\theta}^{*}, \boldsymbol{\phi}^{*}, \boldsymbol{\psi}_{\text{W}}^{*}\}$, the codebook $\mathcal{C}^{*}$, the hybrid beamforming matrices $\mathbf{V}_{\text{A}}^{*}, \mathbf{V}_{\text{D}}^{*}, \mathbf{Q}_{\text{A}}^{*}$ and $\mathbf{Q}_{\text{D}}^{*}$.
   \end{algorithmic}
   \end{algorithm}

\section{Encoder, Decoder, and Codebook Training}
\label{Sec:training}
In this section, we describe the training process in Stage I, where the encoder, decoder, and codebook are optimized. To ensure training stability, we adopt a two-phase training algorithm for the encoder and decoder in this stage. In the first phase, the model is trained without adversarial training, allowing it to first learn the semantic alignment based solely on the $\ell_1$ loss, user-intent relevance loss (LLaVA), and quantization loss. The loss function is given by
\begin{align}
    \mathcal{L}_{\text{phase-I}}(\mathbf{X}_{\text{I}}, \mathbf{Y};\, \mathbf{x}_{\text{T}}) 
    =\ & \mathcal{L}_{1}(\mathbf{X}_{\text{I}}, \mathbf{Y}) 
    + \lambda_{\text{user}}\, \mathcal{L}_{\text{user}}^{\text{LLaVA}}(\mathbf{X}_{\text{I}}, \mathbf{Y};\, \mathbf{x}_{\text{T}}) \notag \\
    & + \lambda_{\text{quant}}\, \mathcal{L}_{\text{quant}}. \label{eq:loss_phase1}
\end{align}
This phase enables the model to learn semantic reconstruction from discrete features, even though the resulting images may lack visual sharpness. After the model has converged and learned a reliable representation space, adversarial training is introduced to improve the perceptual quality and realism of the reconstructed images. In Phase II, the overall loss function is
\begin{align}
    \mathcal{L}_{\text{phase-II}}(\mathbf{X}_{\text{I}}, \mathbf{Y};\, \mathbf{x}_{\text{T}}) 
    = \mathcal{L}_{\text{phase-I}}(\mathbf{X}_{\text{I}}, \mathbf{Y};\, \mathbf{x}_{\text{T}})
+ \lambda_{\text{gen}}\, \mathcal{L}_{\text{gen}}(\mathbf{Y}). \label{eq:loss_phase2}
\end{align}
After updating the generator, the discriminator is trained using the following loss function:
\begin{align}
    \mathcal{L}_{\text{dis}}(\mathbf{X}_{\text{I}},\mathbf{Y}) 
    &= -\mathbb{E}[\log \mathcal{D}_{\text{dis}}(\mathbf{X}_{\text{I}};\boldsymbol{\psi}_{\text{dis}})] \nonumber \\
    &\quad - \mathbb{E}[\log(1 - \mathcal{D}_{\text{dis}}(\mathbf{Y};\boldsymbol{\psi}_{\text{dis}}))], \label{Eq:discriminatorloss}
\end{align}
where the first term encourages high confidence for real images $\mathbf{X}_{\text{I}}$, and the second term penalizes for assigning high confidence to the reconstructed output $\mathbf{Y}$.
In Phase II, the training process alternates between our model and the discriminator. First, we update our model and keep the discriminator to be fixed. Then, we update the discriminator while keeping our model fixed. This two-phase algorithm enables quantized transmission while preserving both the reconstruction fidelity and semantic alignment with the user's intent. The complete training procedure of Stage I is summarized in Algorithm~\ref{alg:stage1}.

\begin{algorithm}[t]
\small
\caption{\small Stage I: Proposed Two-phase Training Algorithm}
\label{alg:stage1}
\begin{algorithmic}[1]
\State \textbf{Input:} Image data and text query in the training datasets; number of epochs for the first and second phases, $N_1$ and $N_2$; learning rate for our model and the discriminator $\eta_{\text{gen}}$ and $\eta_{\text{dis}}$; regularization weights $\lambda_{\text{user}}$, $\lambda_{\text{gen}}$, $\lambda_{\text{quant}}$ and $\gamma$.
\State \textbf{Phase I:}
\For{$i = 1$ \textbf{to} $N_1$} 
    \State Load input image $\mathbf{X}_{\text{I}}$ and query $\mathbf{x}_{\text{T}}$ from the dataset.
    \State Sample the channel response $\mathbf{H}$ and noise $\mathbf{n}$.
    \State Perform forward propagation.
    \State Calculate the loss $\mathcal{L}_{\text{phase-I}}(\mathbf{X}_{\text{I}}, \mathbf{Y};\, \mathbf{x}_{\text{T}})$ based on~\eqref{eq:loss_phase1}.
    \State Update parameters $\{\boldsymbol{\theta}, \boldsymbol{\phi}\}$ and codewords in $\mathcal{C}$ \text{by }$\mathcal{L}_{\text{phase-I}}$.
\EndFor
\State \textbf{Phase II:}
\State Load parameter sets $\{\boldsymbol{\theta}, \boldsymbol{\phi}, \mathcal{C}\}$ trained in the first phase.
\For{$i = 1$ \textbf{to} $N_2$} 
    \State Load input image $\mathbf{X}_{\text{I}}$ and query $\mathbf{x}_{\text{T}}$ from the dataset.
    \State Sample the channel response $\mathbf{H}$ and noise $\mathbf{n}$.
    \State Perform forward propagation.
    \State Calculate the loss $\mathcal{L}_{\text{phase-II}}(\mathbf{X}_{\text{I}}, \mathbf{Y};\, \mathbf{x}_{\text{T}})$ based on~\eqref{eq:loss_phase2}.
    \State Update parameters $\{\boldsymbol{\theta}, \boldsymbol{\phi}\}$ and codewords in $\mathcal{C}$ \text{by }$\mathcal{L}_{\text{phase-II}}$.
    \State Perform forward propagation on discriminator.
    \State Calculate the loss $\mathcal{L}_{\text{dis}}(\mathbf{X}_{\text{I}},\mathbf{Y})$ based on (\ref{Eq:discriminatorloss}).
    \State Update the discriminator parameters $\boldsymbol{\psi}_{\text{dis}}$ \text{by} $\mathcal{L}_{\text{dis}}(\mathbf{X}_{\text{I}},\mathbf{Y})$.
    \EndFor
    \State \textbf{Output}: The optimized parameters $\{\boldsymbol{\theta}^{*}, \boldsymbol{\phi}^{*}\}$ and codebook $\mathcal{C}^{*} := \{ {\mathbf{c}_i^{l}}^{*} \, |\, i = 1, \ldots, N_{\text{cw}},\ l = 1, \ldots, L \}$.
    \end{algorithmic}
\end{algorithm}

\section{Semantic-Aware Hybrid Beamforming Design}
\label{Sec:beamforming}

In this section, we describe the optimization procedure in Stage~II, which trains the weight module to assign importance weights to subcarriers in the same transmission block and utilizes these weights to design the hybrid beamforming matrices. The objective of the weight module is to assign higher weights to subcarriers that transmit features which are more important in order to reduce the total loss $\mathcal{L}_{\text{total}}$. That is, the hybrid beamforming module allocates more resources to those features which are more important. We formulate the weighting mechanism as a contextual bandit problem. Based on the weight matrix given by the weight module in each round, we propose the WMMSE HBF algorithm to determine the hybrid beamforming matrices. When the weight module is being trained, the encoder, decoder, and codebook are fixed using parameters obtained from Stage~I.

\subsection{Contextual Bandit Formulation for SA-HBF}
\begin{figure}[t]
    \centering
    \includegraphics[width=\linewidth]{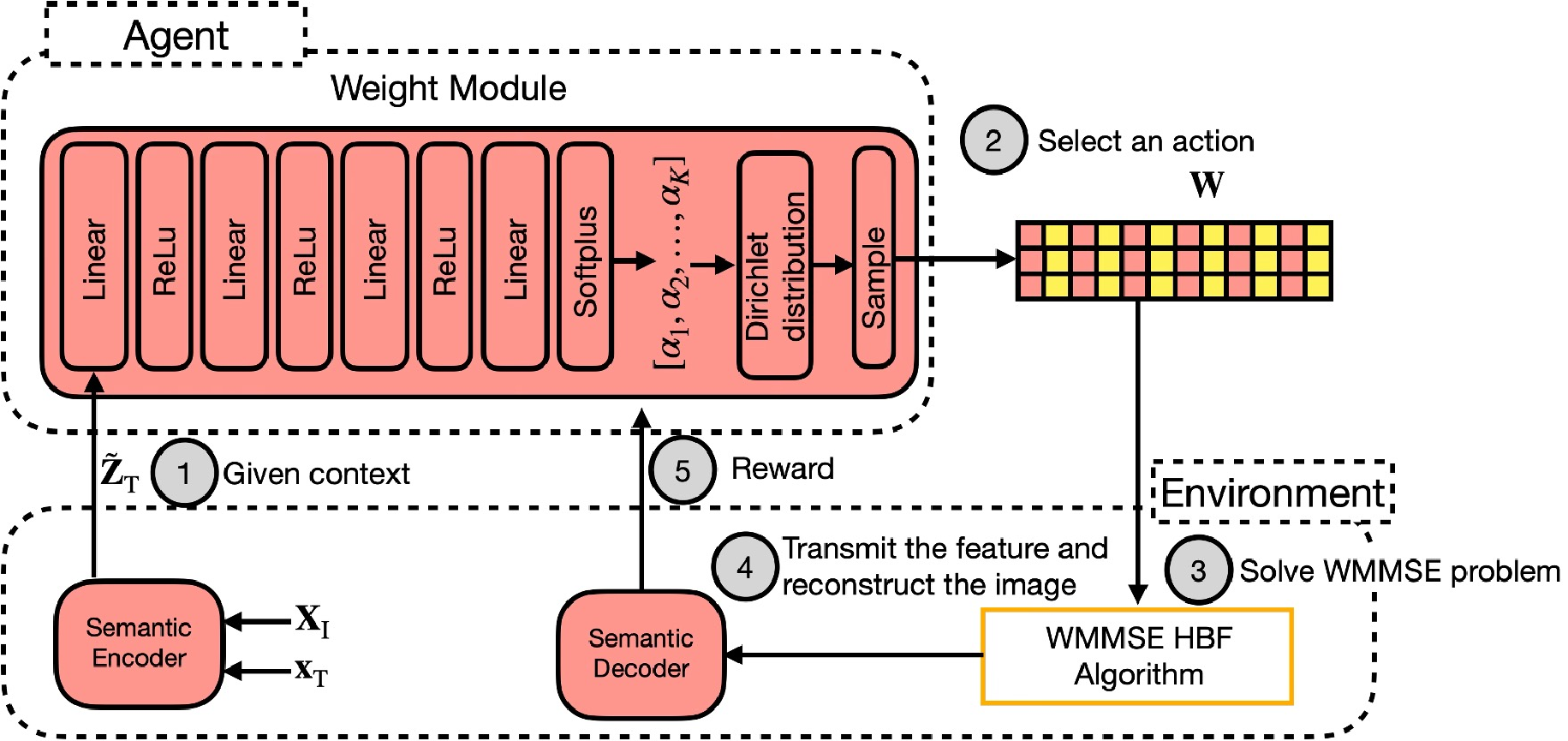}
    \caption{The contextual bandit framework for semantic-aware weight assignment.}
    \label{fig:bandit}
\end{figure}
Contextual bandit~\cite{Latt2020} is a type of reinforcement learning algorithm. In particular, an agent observes the context, selects an action, and receives a reward based on the outcome of the selected action in each round. The goal is to learn a policy that minimizes the regret function over multiple rounds. Each round is modeled as an independent decision-making instance. As depicted in Fig.~\ref{fig:bandit}, we treat the transmission of an image as a single round in the contextual bandit problem. In each round $t$, the input image $\mathbf{X}_{\text{I}}^{(t)}$ and text query $\mathbf{x}_{\text{T}}^{(t)}$ are passed through the encoder to produce semantic features $\mathbf{\tilde{Z}}_{\text{T}}^{(t)}$. The feature matrix $\mathbf{\tilde{Z}}_{\text{T}}^{(t)}$ serves as the context in round $t$. In the contextual bandit framework, we denote the context as $\mathbf{c}^{(t)}$. The weight module then serves as an agent in the contextual bandit framework. It processes the feature matrix $\mathbf{\tilde{Z}}_{\text{T}}^{(t)}$ in a row by row manner. For each transmission block $b$, it takes the $b$th row vector $\mathbf{\tilde{z}}_{\text{T},b}^{(t)}$ as input and outputs the corresponding weight vector $\mathbf{w}_{b}^{(t)}$, as shown in equation~\eqref{eq:weight}. After all $B$ transmission blocks have been processed, the resulting weight vectors are stacked to form the weight matrix $\mathbf{W}^{(t)}$. This matrix represents the action in round $t$, denoted as $\mathbf{a}^{(t)}$, and is sampled according to the policy $\pi_{\boldsymbol{\psi}_{\text{W}}}$, which is parameterized by $\boldsymbol{\psi}_{\text{W}}$. That is, the action corresponds to the weight matrix $\mathbf{W}^{(t)}$ in QO-ISC. This weight matrix serves as the input for the WMMSE algorithm to design the HBF matrices. Subsequently, the semantic features are transmitted over the wireless channel. Upon reception, the receiver reconstructs the image $\mathbf{Y}^{(t)}$ and generates a reward $r^{(t)}$, defined as the negative value of total loss: $r^{(t)} = -\mathcal{L}_{\text{total}}(\mathbf{X}_{\text{I}}^{(t)}, \mathbf{Y}^{(t)};\, \mathbf{x}_{\text{T}}^{(t)})$. Within this contextual bandit framework, the encoder, decoder, codebooks, and the physical wireless channel constitute the environment. We define a context space $\mathcal{X}$ which includes all the possible contexts that the agent can observe from the environment as well as an action space $\mathcal{A}$ that includes all the possible weight matrices produced by the weight module. The contextual bandit problem runs over $T$ rounds. The policy $\pi: \mathcal{X} \rightarrow \mathcal{A}$ maps each context $\mathbf{c} \in \mathcal{X}$ to an action $\mathbf{a} \in \mathcal{A}$. The agent's goal is to learn a policy that minimizes the regret over $T$ rounds, where the regret is defined as
\begin{align}
    \mathcal{R}(T) = \sum_{t=1}^{T} \big[ & \ {r}^{(t)*}(\mathbf{c}^{(t)}, \mathbf{a}^{(t)}) - {r}^{(t)}(\mathbf{c}^{(t)}, \mathbf{a}^{(t)}) \big],
\end{align}
where ${r}^{(t)*}$ is the reward of the optimal policy in round $t$, and ${r}^{(t)}$ is the reward of the current policy. A lower regret indicates that the learned policy performs closer to the optimal one. 

In bandit problems, the agent makes a sequence of decisions. The main challenge lies in the exploration-exploitation dilemma. One of the widely used methods that can balance the exploration and exploitation tradeoff in bandit problems is the Thompson sampling method~\cite{Agrawal2013}. In each round, Thompson sampling maintains a posterior distribution over the policy parameters and samples from this distribution to select an action. The posterior distribution is then updated after the reward from the action is observed.

Since the weight vector $\mathbf{w}_{b} = [w_{b,1} \cdots w_{b,K}] \in \mathbb{R}^{1 \times K }$ lies in the $K$-dimensional probability simplex, i.e. $w_{b,k} \in [0,1]$ and $\sum_{k=1}^{K} w_{b,k} = 1$, we model it using a Dirichlet distribution~\cite{Samuel2000}. The Dirichlet distribution is parameterized by a vector of positive concentration parameters $\boldsymbol{\alpha}^{(t)} = (\alpha_1^{(t)}, \ldots, \alpha_K^{(t)}) \in \mathbb{R}^K_{++}$. We adopt a neural network to model the posterior over weight vectors by generating the parameters of a Dirichlet distribution conditioned on the input context, similar to variational methods used in Bayesian deep bandits~\cite{Carlos2018}. Specifically, as illustrated in Fig.~\ref{fig:bandit}, the weight module observes the feature matrix $\mathbf{\tilde{Z}}_{\text{T}}^{(t)}$ from the environment. Each row of the feature matrix is passed through a series of fully connected layers, each followed by a ReLU activation function. To ensure that the concentration parameters $\boldsymbol{\alpha}^{(t)}$ are strictly positive, we apply a softplus activation function at the output layer. The resulting vector $\boldsymbol{\alpha}^{(t)}$ parameterizes the Dirichlet distribution, from which we sample the weight vector $\mathbf{w}_b^{(t)} \sim \operatorname{Dir}(\boldsymbol{\alpha}^{(t)})$. The sampled action $\mathbf{w}_b^{(t)}$ is stacked to obtain the matrix $\mathbf{W}^{(t)}$ which is used to compute the hybrid beamforming matrices. The receiver reconstructs the image. The reward ${r}^{(t)}$ is then computed based on the reconstructed image and the original image. Based on the reward ${r}^{(t)}$, we update the policy parameters $\pi_{\boldsymbol{\psi}_{\text{W}}}$ using the policy gradient method~\cite{Ronald1992}. Specifically, we maximize the expected reward
\begin{align}
    J(\pi_{\boldsymbol{\psi}_{\text{W}}}) 
    = \mathbb{E}_{\mathbf{c} \sim \mathcal{X}}\, 
    \mathbb{E}_{\mathbf{a}\sim \pi_{\boldsymbol{\psi}_{\text{W}}}(\cdot \mid \mathbf{c})}
    \left[ r(\mathbf{c}, \mathbf{a}) \right].
\end{align}
The gradient of this objective can be written as
\begin{align}
    \nabla_{\boldsymbol{\psi}_{\text{W}}} J(\pi_{\boldsymbol{\psi}_{\text{W}}})
    = \mathbb{E}_{\mathbf{c},\, \mathbf{a}} \big[ r(\mathbf{c}, \mathbf{a})
    \nabla_{\boldsymbol{\psi}_{\text{W}}} \log 
    \pi_{\boldsymbol{\psi}_{\text{W}}}(  \mathbf{a} \mid \mathbf{c})\big].
    \label{eq:gradientloss}
\end{align}
We estimate the gradient in \eqref{eq:gradientloss} using a stochastic sample in each round, which is expressed as
\begin{align}
\nabla_{\boldsymbol{\psi}_{\text{W}}} J(\pi_{\boldsymbol{\psi}_{\text{W}}}) \approx r^{(t)}(\mathbf{c}^{(t)}, \mathbf{a}^{(t)}) \nabla_{\boldsymbol{\psi}_{\text{W}}} \log \pi_{\boldsymbol{\psi}_{\text{W}}}(\mathbf{a}^{(t)} \mid \mathbf{c}^{(t)}). \label{eq:stochasitcgradientloss}
\end{align}
This gradient is used to update $\boldsymbol{\psi}_{\text{W}}$ via stochastic gradient ascent. 
The algorithm of Stage II is summarized in Algorithm~\ref{alg:stage2}.

\begin{algorithm}[t]
    \small
    \caption{\small Stage II: Contextual Bandit-based SA-HBF Algorithm}
    \label{alg:stage2}
    \begin{algorithmic}[1]
    \State \textbf{Input:} Image data and text query in the training datasets; the number of rounds $T$; the learning rate $\eta_{\text{W}}$ for the weight module.
    \State Randomly Initialize the weight module $\boldsymbol{\psi}_{\text{W}}$.
    \For{$t = 1$ \textbf{to} $T$} 
        \State Load input image $\mathbf{X}_{\text{I}}^{(t)}$ and text query $\mathbf{x}_{\text{T}}^{(t)}$ from the dataset.
        \State Sample the channel response $\mathbf{H}$ and noise $\mathbf{n}$.
        \State Perform forward pass to obtain the context $\mathbf{c}^{(t)} := \mathbf{\tilde{Z}}_{\text{T}}^{(t)}$.
        \State Pass row vector $\mathbf{\tilde{z}}_{\text{T},b}^{(t)}$ of $\mathbf{\tilde{Z}}_{\text{T}}^{(t)}$ through the weight module to generate Dirichlet parameters $\boldsymbol{\alpha}^{(t)}$.
        \State Sample the weight vector $\mathbf{w}_{b}^{(t)}$ from the Dirichlet distribution.
        \State Stack them to obtain action $\mathbf{a}^{(t)} := \mathbf{W}^{(t)}$.
        \For{$b = 1$ \textbf{to} $B$} 
            \State Given weight vector $\mathbf{w}_{b}^{(t)}$, obtain the optimal hybrid precoder $\mathbf{V}_{\text{A}}^{*}$, $\mathbf{V}_{\text{D}}^{*}$ and hybrid combiner $\mathbf{Q}_{\text{A}}^{*}$, $\mathbf{Q}_{\text{D}}^{*}$ from Algorithm~\ref{alg:mmseHBF}.
            \State Transmit the quantized feature matrix $\mathbf{\bar{Z}}_{\text{T}, b}$ for $b$th transmission block through the wireless channel.
        \EndFor
        \State Stack the received data symbol to form $\mathbf{\bar{Z}}_{\text{Y}}$ and reconstruct the image $\mathbf{Y}^{(t)}$ at the receiver.
        \State Calculate the reward $r^{(t)}:=-\mathcal{L}_{\text{total}}(\mathbf{X}_{\text{I}}^{(t)}, \mathbf{Y}^{(t)};\, \mathbf{x}_{\text{T}}^{(t)})$.
        \State Update the weight module parameters $\boldsymbol{\psi}_{\text{W}}$ by using gradient ascent based on the gradient in~\eqref{eq:stochasitcgradientloss}.
    \EndFor
    \State \textbf{Output}: The optimized policy $\pi_{\boldsymbol{\psi}_{\text{W}}}^{*}$.
    \end{algorithmic}
\end{algorithm}

\subsection{WMMSE Hybrid Beamforming Algorithm}
In each round $t$ of the contextual bandit problem, the weight module outputs a weight matrix $\mathbf{W}$, which consists of weight vectors $\mathbf{w}_{b} = [w_{b, 1} \cdots w_{b,K}]$ corresponding to each transmission block $b = 1, \ldots, B$. The weight vector $\mathbf{w}_{b}$ is used to compute the hybrid beamforming matrices. For a given weight vector $\mathbf{w}_b$ for transmission block $b$, we formulate the following WMMSE problem to minimize the distortion of data symbols
\begin{align}
    \underset{\mathbf{V}_{\text{A}},\, \mathbf{V}_{\text{D}},\, \mathbf{Q}_{\text{A}},\, \mathbf{Q}_{\text{D}}, \, \boldsymbol{\beta}}{\text{minimize}} \quad &
    \sum_{k=1}^{K} w_{b,k} \, \mathbb{E} \left[
        \left\| \beta_{k}^{-1} \hat{\mathbf{s}}_{k} - \mathbf{s}_{k} \right\|^{2}
    \right] \label{eq:P2a} \\
    \text{subject to constr}&\text{aints}
    ~\eqref{eq:P0_constraint1},\ \eqref{eq:P0_constraint2},\ \text{and } \eqref{eq:P0_constraint3}, \nonumber
\end{align}
where $\beta_{k}$ is a normalizing factor in subcarrier $k$. $\beta_{k}$ is introduced in~\cite{Lin2019} in order to decouple the hybrid precoding and combining subproblems, account for noise, and simplify the transmit power constraint. We define the vector of these normalizing factors across all subcarriers as $\boldsymbol{\beta} = [\beta_{1}, \dots, \beta_{K}]$. To solve the WMMSE problem in \eqref{eq:P2a}, we extend the HBF MMSE formulation in~\cite{Lin2019}. When the weights are the same, i.e. $w_{b,k} = 1/K$ for all $k$, the formulation reduces to that of~\cite{Lin2019}. However, when the weights are not the same, two key distinctions arise:
\begin{enumerate}
    \item \textit{Per-subcarrier weight allocation:} Each subcarrier is assigned a different weight, allowing the system to adaptively allocate more transmission resources to those subcarriers carrying features that are semantically important.
    \item \textit{Contextual bandit-based weight update:} The weights are optimized through a contextual bandit framework, using the negative value of the total loss $-\mathcal{L}_{\text{total}}$ as the reward. 
\end{enumerate}
Problem~\eqref{eq:P2a} is solved by alternating optimization, where the hybrid combiners $\mathbf{Q}_{\text{D}}$ and $\mathbf{Q}_{\text{A}}$ are fixed at first. We optimize the hybrid precoders $\mathbf{V}_{\text{D}}$ and $\mathbf{V}_{\text{A}}$. We then fix the hybrid precoders $\mathbf{V}_{\text{D}}$ and $\mathbf{V}_{\text{A}}$ as constants while optimizing the hybrid combiners $\mathbf{Q}_{\text{D}}$ and $\mathbf{Q}_{\text{A}}$. This alternating minimization strategy is repeated until convergence. The detailed process is described in the Appendix.

\textbf{\textit{Remark 1:}} In our proposed QO-ISC framework, the semantic importance weights $\mathbf{W}$ serve as a cross-layer bridge between semantic coding and the physical layer. Unlike traditional hybrid beamforming, which assigns the weight based on physical layer metrics such as channel quality to maximize the sum rate, our proposed QO-ISC framework leverages semantic importance to design the hybrid beamforming matrices. 

\subsection{Computational and Training Overhead}
The training procedure has three stages. In Stage~I, we train the encoder, decoder, and codebook while keeping the CLIP backbone to be frozen. Since only the projection MLPs, FiLM layers, convolutional decoder, and codebooks are being updated, the number of trainable parameters is much smaller than that of CLIP and the cost is comparable to a learned image codec with a frozen backbone. In Stage~II, we train only the weight module, with the semantic encoder, decoder, and codebook remain fixed. This involves processing the feature matrix $\mathbf{\tilde{Z}}_{\text{T}}$ with a fully connected network in each round. The computational cost per round for the weight module scales as $\mathcal{O}(B K N_{\text{L}})$. Stage~III fine-tunes these small modules, while the CLIP backbone remains fixed, resulting in a complexity similar to Stage~I. In WMMSE-based hybrid beamforming algorithm, given a fixed weight vector $\mathbf{w}_b$ for transmission block $b$, it alternates between precoder and combiner updates. The complexity per iteration is 
$
\mathcal{O}\big(K (N_\mathrm{t} N_\mathrm{RF}^2 N_\mathrm{S}+ N_\mathrm{r} N_\mathrm{RF}^2)\big).
$

\begin{figure*}[t]
    \centering
    \includegraphics[width=\linewidth]{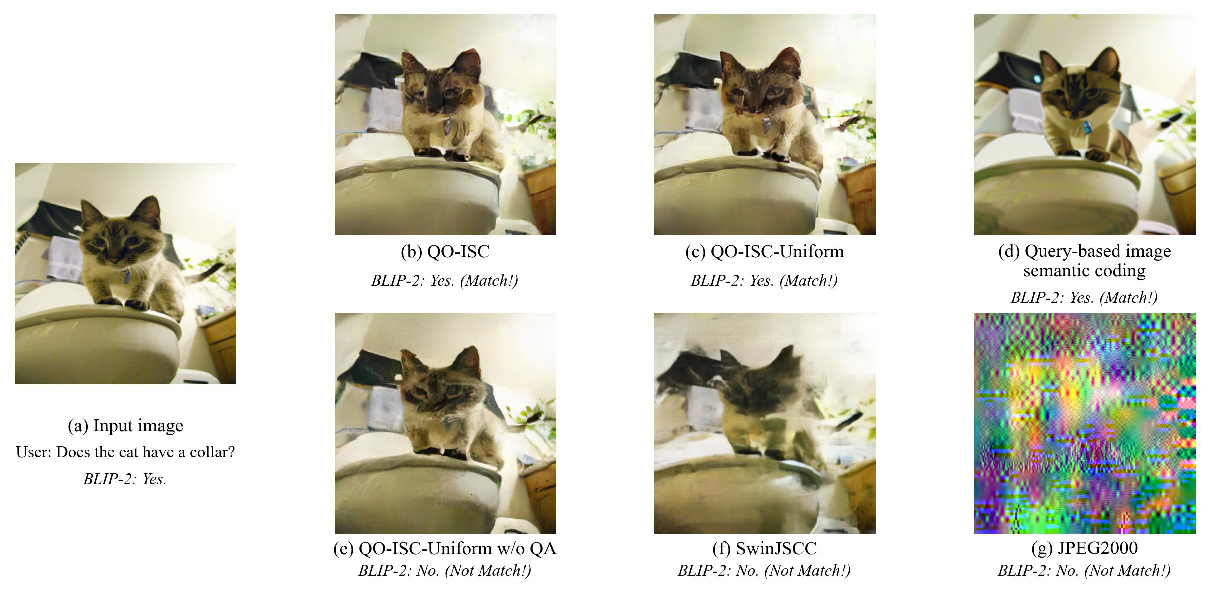}
    \caption{\small Visualization of the images reconstructed by our proposed QO-ISC and baseline methods with the SNR sets to $-22$ dB. The text below each image shows the answer generated by the BLIP-2 model to the query ``Does the cat have a collar?''. The italicized text, \textit{Match!} or \textit{Not Match!}, explicitly indicates whether BLIP-2's answer aligns with the ground truth answer.}
    \label{fig:visual}
\end{figure*}

\section{Performance Evaluation}
\label{Sec:simulation}
In this section, we evaluate the performance of the proposed QO-ISC framework and compare it with several baseline models. The QO-ISC framework employs the pretrained CLIP ViT-B/16\footnote{Model available at https://github.com/openai/CLIP} model with a patch size of 16 and a projection dimension of $D = 512$. In the encoder, we select the transformer layers $\mathcal{M}_{\text{S}} = \{3, 6, 9, 11\}$ from a total of $M = 12$ layers to capture a multi-scale semantic representation from low-level details to high-level abstractions. Consequently, the decoder utilizes $N = 4$ convolutional layers. Each feature vector is partitioned into segments of length $N_{\text{L}} = 64$, leading to $L = \frac{D}{N_{\text{L}}} = 8$ segments per vector. Each segment is quantized using a codebook containing $N_{\text{cw}} = 64$ codewords. The parameters  $\lambda_{\text{user}}$, $\lambda_{\text{gen}}$, $\lambda_{\text{quant}}$ and $\gamma$  are set to 1.5, 0.5, 0.5, and 1.2, respectively. We use low density parity check (LDPC) codes with rate $1/2$ and 8-ary phase shift keying modulation. We utilize nanoLLaVA\footnote{Model available at \url{https://huggingface.co/qnguyen3/nanoLLaVA}} to determine the user-intent relevance loss (LLaVA) $\mathcal{L}_{\text{user}}^{\text{LLaVA}}$. For the SA-HBF configuration, the numbers of transmit and receive antennas are set to $N_{\text{t}} = N_{\text{r}} = 64$, respectively. We consider $N_{\text{RF}} = 2$ RF chains, $K = 64$ subcarriers, and $N_{\text{S}} = 2$ data streams per subcarrier. In the channel model, we assume there are $N_{\text{C}} = 5$ clusters and $N_{\text{R}} = 10$ rays per cluster. The angles of arrival and departure, $\theta_{c,r}^{\text{r}}$ and $\theta_{c,r}^{\text{t}}$, are generated according to the Laplacian distribution, where the mean cluster angles are uniformly distributed in $[0, 2\pi]$. The generalized QO-ISC framework is trained on the VQA dataset~\cite{Stan2015}. We compare its performance with the following baseline models:
\begin{itemize}
    \item \textbf{QO-ISC-Uniform:} This model uses the same architecture as QO-ISC but all features have the uniform weight in SA-HBF.
    \item \textbf{QO-ISC-Uniform without (w/o) query alignment (QA):} This model adopts the same architecture as QO-ISC but does not include the weight module and feature alignment with text query. It assigns the same weight to all subcarriers and does not consider user-intent relevance loss in training.
    \item \textbf{SwinJSCC-large~\cite{Yang2025}:} This is a state-of-the-art JSCC codec built on the Swin Transformer backbone and does not incorporate any query-based alignment mechanisms or pretrained parameters.
    \item \textbf{Query-based image semantic coding~\cite{chen2024}}: This model transmits the answer generated by an LLM along with a compressed image to the receiver, and employs a pretrained diffusion model to reconstruct the image based on the generated answer and the compressed image.
    \item \textbf{Traditional image coding (JPEG2000)~\cite{JPEG2000}:} As a conventional baseline, JPEG2000 is used to compress and reconstruct images at the pixel level, without considering semantic relevance.
\end{itemize}
Since the proposed QO-ISC framework is optimized by using $\mathcal{L}_{\text{user}}^{\text{LLaVA}}$ based on LLaVA, comparing its performance with other baselines through the same model may lead to biased results. Thus, we utilize BLIP-2~\cite{Li2022} as an independent evaluator which is not used in any part of our training pipeline. Specifically, we use BLIP-2 to generate answers from both the original and reconstructed images under the same query. The prompt provided to BLIP-2 is ``Question: \{question\} Answer:", where \{question\} denotes the text query. For a given image and prompt, we use greedy decoding~\cite{holtzman2019curious} to generate a deterministic answer. We separate the questions into different question types, where question types are identified by keyword rules applied to the question text. Each generated answer is normalized by converting it to lowercase, removing punctuation and articles, and collapsing whitespace. The comparison is then performed on each question type. For counting questions, the numerical values extracted from the two answers are compared. For color, location, and yes/no questions, the corresponding keywords are extracted from the two answers based on a predefined term list. The answers are considered to be a match if the two answers share at least one common term. For open-ended questions that do not fall into these categories, we assess the semantic similarity between the two answers using the BERTScore~\cite{Zhang2020}, which measures the similarity of two texts by comparing their token embeddings derived from a pretrained BERT~\cite{Devlin2018} model. To evaluate the semantic consistency across all question types, we also use the \textit{answer match rate} as a metric. This metric averages a binary match indicator for counting, color, location, and yes/no questions, together with the BERTScore for open-ended questions.

While predicting the same answer does not necessarily mean the model has the same confidence, we further introduce the loss function $\mathcal{L}_{\text{user}}^{\text{BLIP-2}}$ to quantify how well the reconstructed image preserves the query-relevant semantic content. Similar to $\mathcal{L}_{\text{user}}^{\text{LLaVA}}$ in~(\ref{eq:loss_user}), $\mathcal{L}_{\text{user}}^{\text{BLIP-2}}$ is defined as the cosine distance between the multimodal embeddings of the original and reconstructed images conditioned on the same query. The only difference with $\mathcal{L}_{\text{user}}^{\text{LLaVA}}$ is that the multimodal embeddings are extracted by the frozen BLIP-2 model. A lower $\mathcal{L}_{\text{user}}^{\text{BLIP-2}}$ indicates better preservation of the query-relevant semantic content in the reconstructed image. The reported performance metrics are averaged across the results generated from various channel realizations.

\subsection{Performance Comparison}
We evaluate the generalization capability of the proposed QO-ISC framework. We train our proposed QO-ISC on a subset of the VQA dataset by excluding animal-related samples and perform zero-shot evaluation on a subset containing only animal-related samples. Fig.~\ref{fig:visual} shows the reconstructed images generated by QO-ISC and baseline models with the SNR sets to $-22$ dB. Compared with the baselines, QO-ISC better preserves semantic features related to the user's prompt, as it extracts features related to the text query, which is the cat's collar, and prioritizes them during wireless transmission. When using uniform weight in the SA-HBF, critical features are no longer prioritized during transmission, which makes the cat's collar slightly blurry. The query-based image semantic coding in~\cite{chen2024} leverages the answer generated by LLM to guide the diffusion model at the decoder. Although the reconstructed image depicts a collar on the cat, it differs significantly from the one in the original image because the diffusion model hallucinates the details based on the generated answer. Models with uniform weight and without query alignment overlook the query-related features and instead preserve the general content such as the cat, rather than user-intent content such as the collar. SwinJSCC overfits the training distribution and fails to generalize to unseen animal dataset because it lacks the mechanism to prioritize query-specific features and the frozen LAM pretrained backbone to improve generalization. JPEG2000 exhibits severe visual artifacts under low SNR because its pixel-wise compressed bitstream is corrupted.
\begin{figure}[t]
    \centering
    
    \begin{subfigure}[b]{0.235\textwidth}
        \centering
        \includegraphics[width=\textwidth]{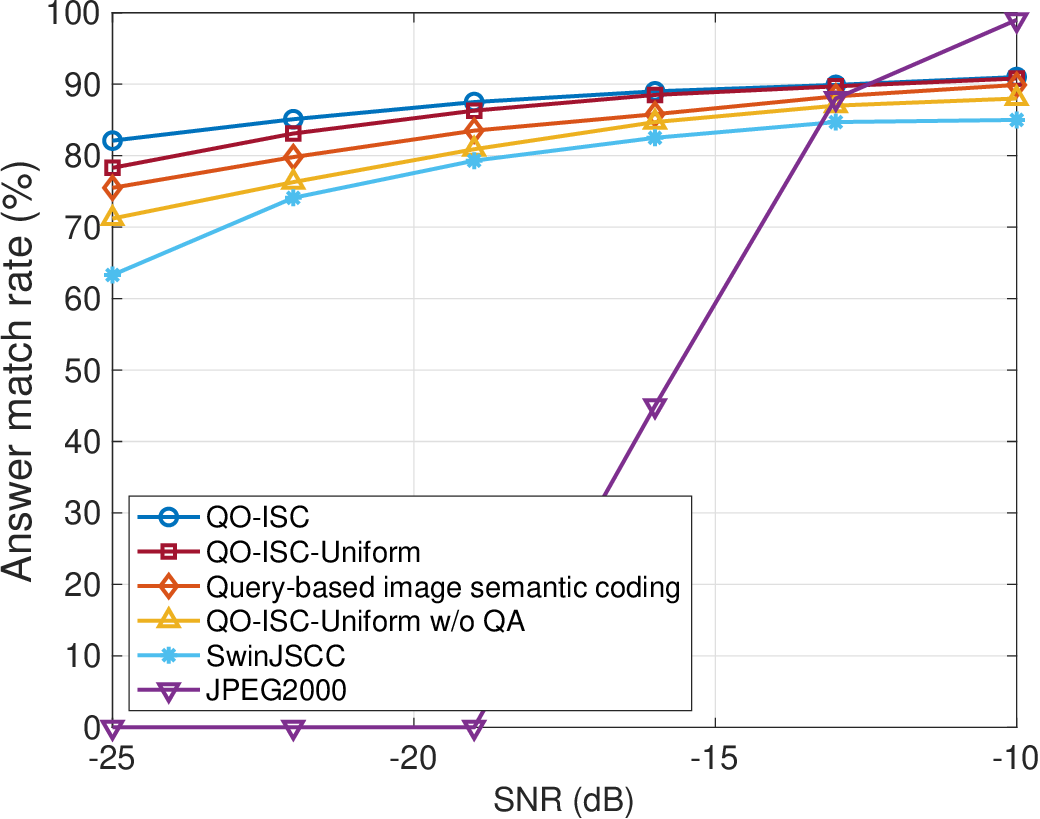}
        \caption{}
        \label{fig:suba}
    \end{subfigure}
    \hfill
    \begin{subfigure}[b]{0.235\textwidth}
        \centering
        \includegraphics[width=\textwidth]{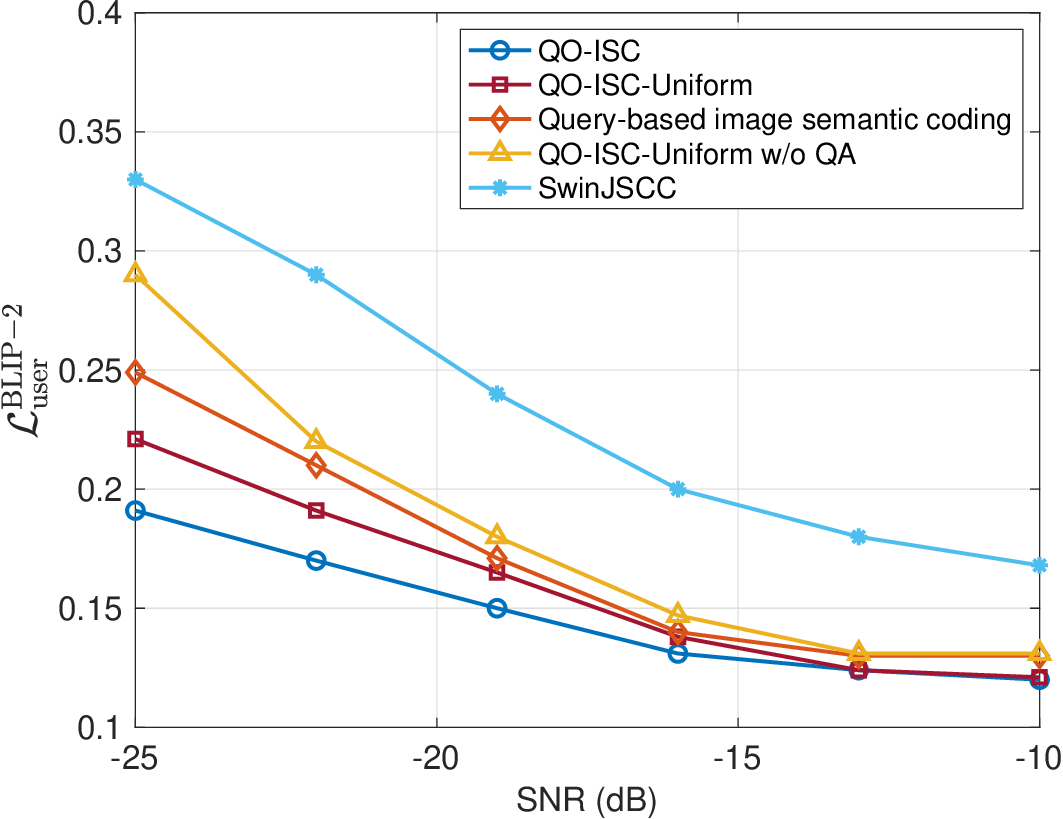}
        \caption{}
        \label{fig:subb}
    \end{subfigure}
    \begin{subfigure}[b]{0.235\textwidth}
        \centering
        \includegraphics[width=\textwidth]{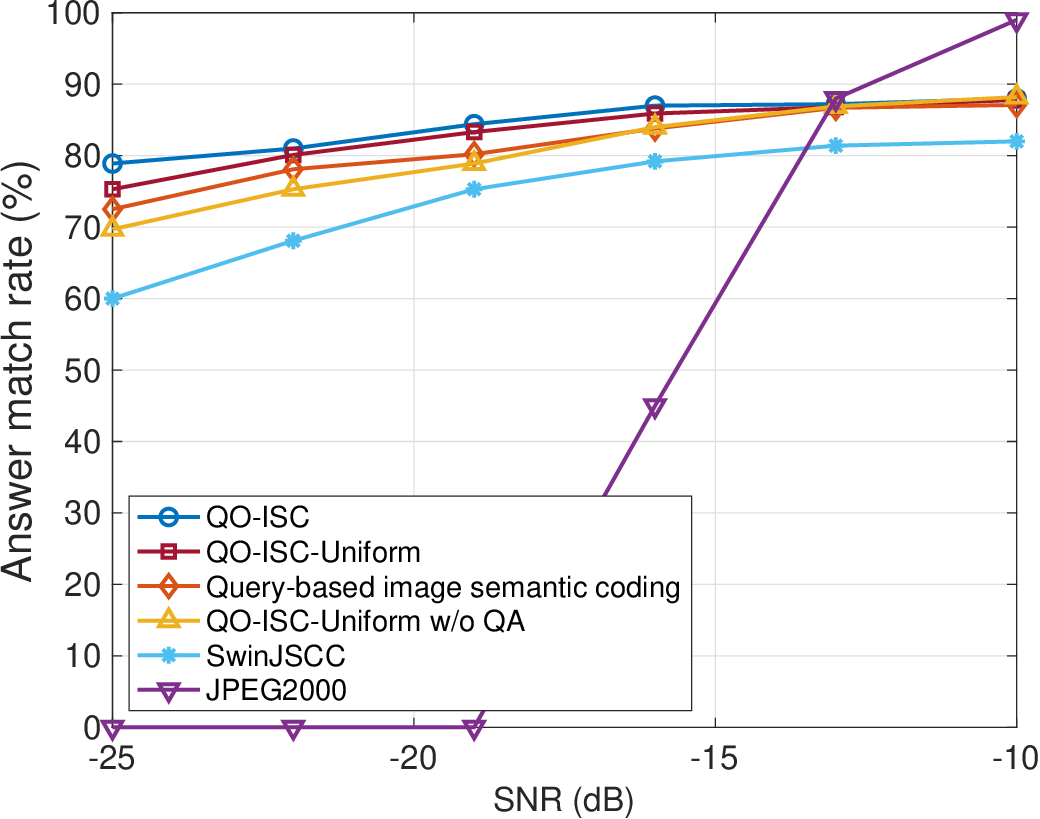}
        \caption{}
        \label{fig:suba}
    \end{subfigure}
    \hfill
    \begin{subfigure}[b]{0.235\textwidth}
        \centering
        \includegraphics[width=\textwidth]{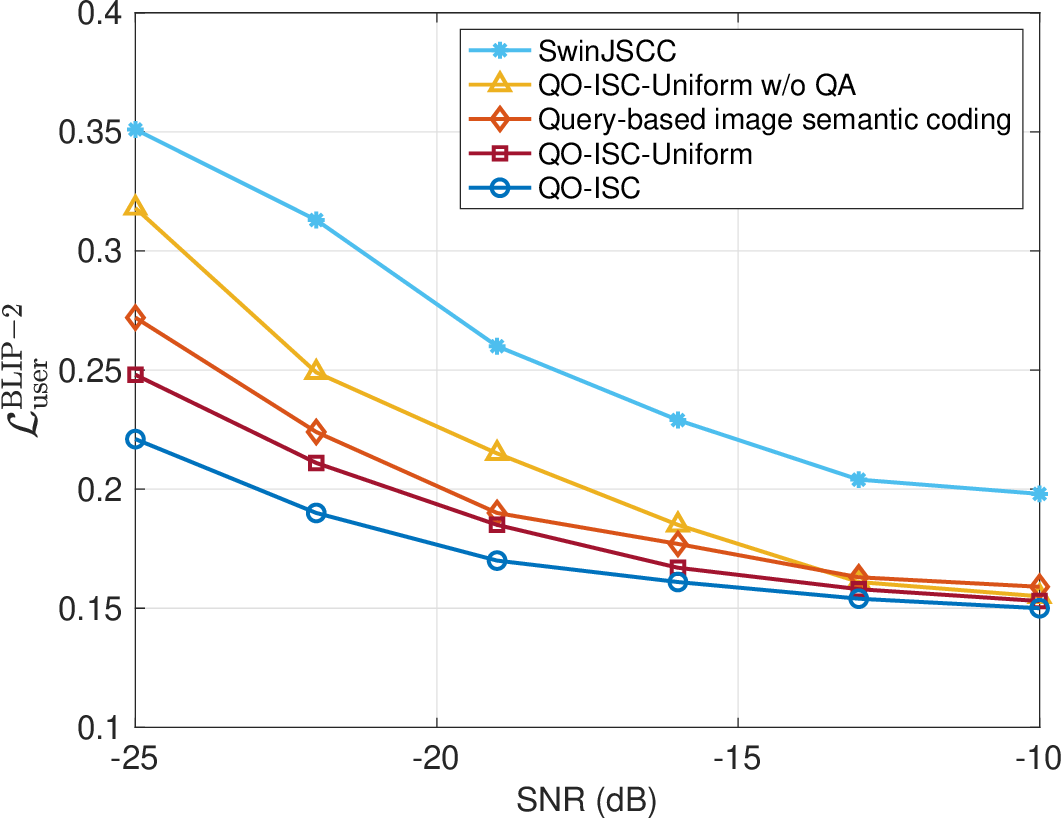}
        \caption{}
        \label{fig:subb}
    \end{subfigure}
    \caption{The answer match rate and user-intent relevance loss (BLIP-2) $\mathcal{L}_{\text{user}}^{\text{BLIP-2}}$ versus the SNR. (a) and (b) show the results for training on the non-animal subset and testing on the animal subset. (c) and (d) show the results for training on the non-human subset and testing on the human subset.}
    \label{fig:general}
\end{figure}

Simulation results in Figs.~\ref{fig:general}(a) and (b) show the answer match rate and the user-intent relevance loss (BLIP-2) $\mathcal{L}_{\text{user}}^{\text{BLIP-2}}$ versus the SNR. Our proposed generalized QO-ISC achieves the lowest $\mathcal{L}_{\text{user}}^{\text{BLIP-2}}$ when the SNR is between $-25$ dB and $-10$ dB. At high SNR, our proposed generalized QO-ISC performs similarly to the QO-ISC-Uniform, QO-ISC-Uniform without QA, and query-based image semantic coding baselines in terms of answer match rate. This occurs because most data can reliably be transmitted, leading the models to transmit all image details regardless of their query relevance. Since JPEG2000 transmits pixel-level information, the image can almost be perfectly reconstructed at high SNR. However, its performance degrades significantly at low SNR. At low SNR, the effectiveness of the proposed generalized QO-ISC becomes apparent. Specifically, at an SNR of $-25$ dB, by employing the importance weighting mechanism   in the SA-HBF module, our framework prevents severe degradation and improves the answer match rate by 4.8\%. Furthermore, by incorporating query alignment, the model prioritizes features relevant to the user's intent, yielding a 9\% improvement in answer match rate compared to QO-ISC-Uniform without QA. When compared to the query-based semantic coding baseline~\cite{chen2024}, our method achieves an 8\% gain. When evaluating the answer match rate of SwinJSCC, the performance is improved by 7.6\% at $\text{SNR} =-10$ dB and 30\% at $\text{SNR}=-25$ dB due to the use of the frozen LAM. We also evaluate the generalization performance in Figs.~\ref{fig:general}(c) and (d), where the models are trained on a subset excluding the human-related samples and are tested on a subset containing only the human-related samples. Our proposed generalized QO-ISC framework demonstrates better generalization capability.

\subsection{Impact of Number of Antennas and SA-HBF}

\begin{figure}[t]
    \centering
    \includegraphics[width=0.7\linewidth]{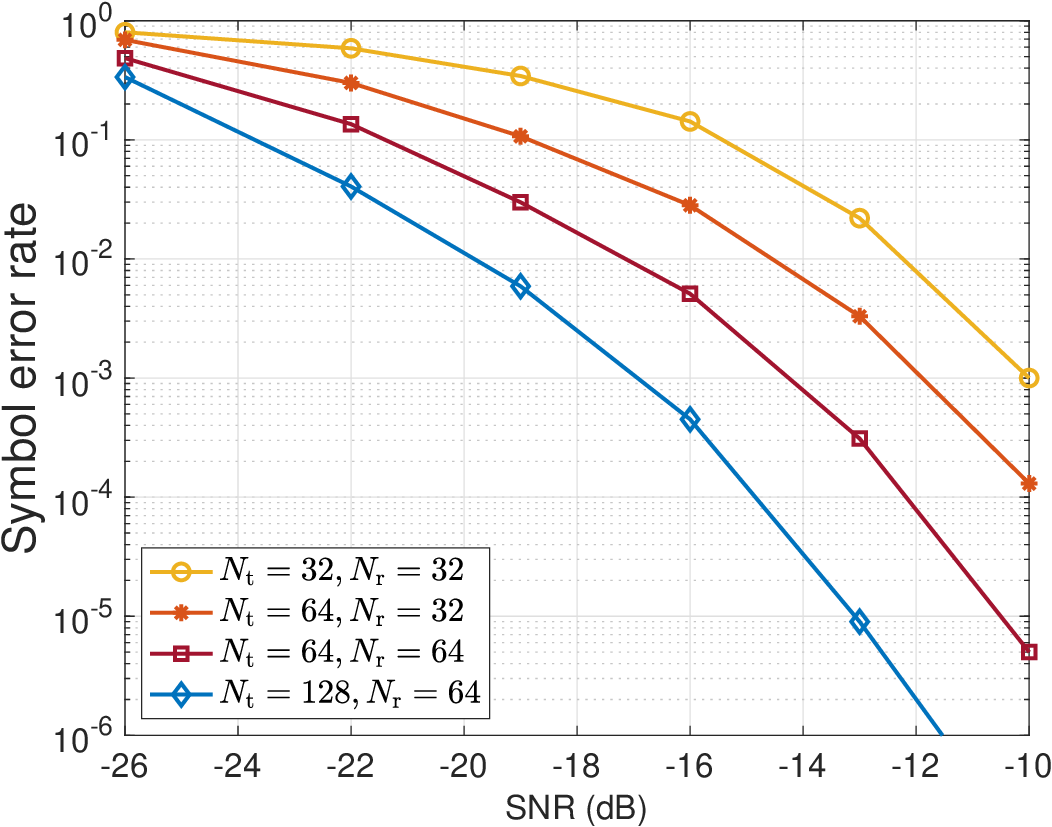}
    \caption{Symbol error rate versus the SNR.}
    \label{fig:Symbolerrorrate}
\end{figure}

\begin{figure}[t]
  \centering
   \begin{subfigure}[b]{0.235\textwidth}
      \centering
      \includegraphics[width=\linewidth]{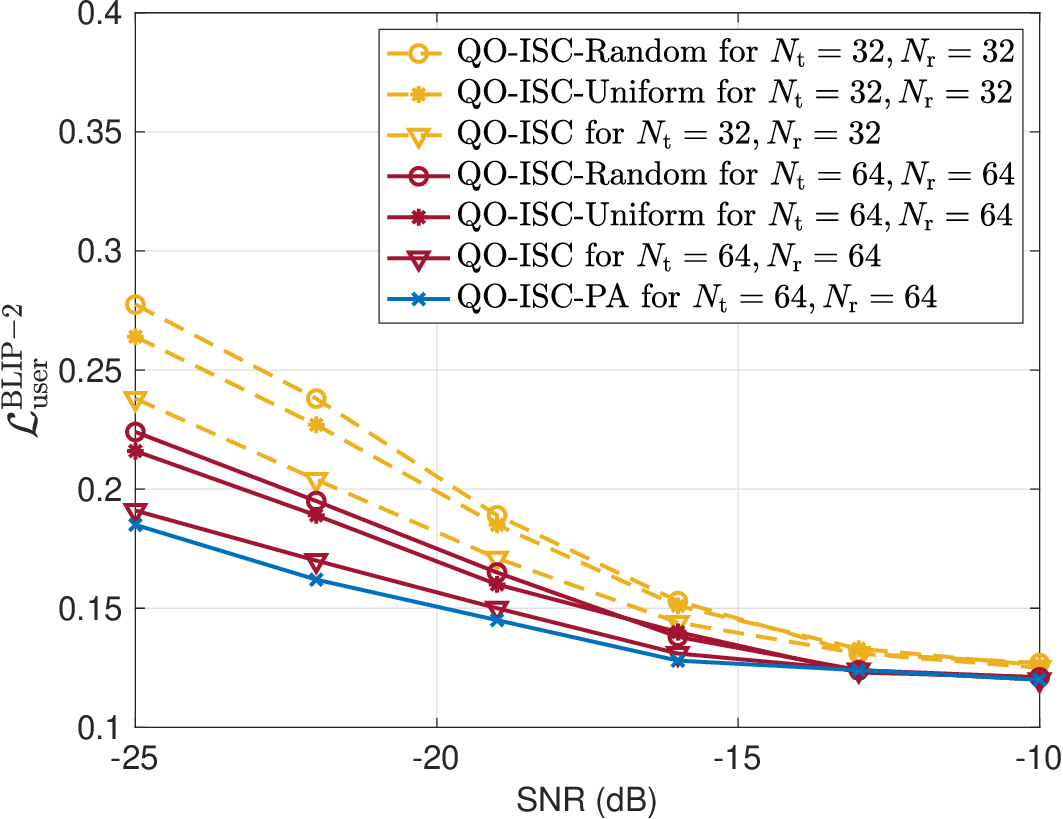}
       \caption{}
   \end{subfigure}
   \hfill
   \begin{subfigure}[b]{0.235\textwidth}
       \centering
       \includegraphics[width=\linewidth]{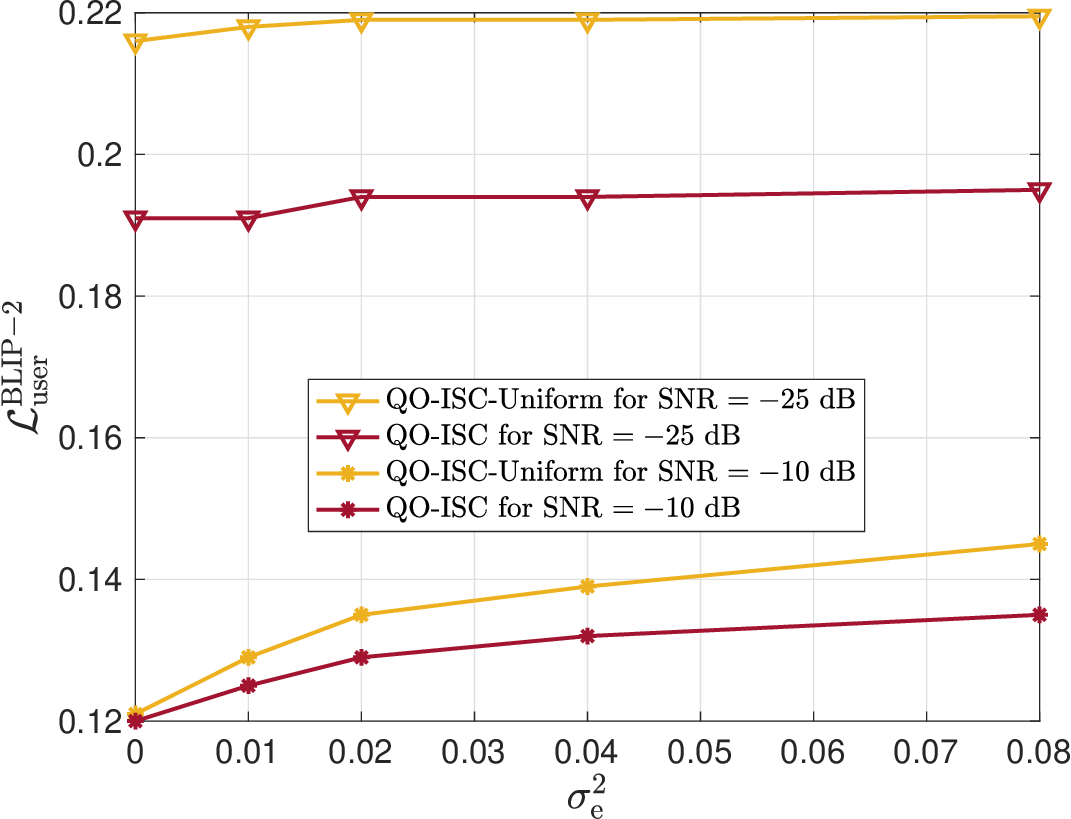}
       \caption{}
   \end{subfigure}
  \caption{The user-intent relevance loss (BLIP-2) $\mathcal{L}_{\text{user}}^{\text{BLIP-2}}$ versus (a) SNR and (b) channel estimation error variance $\sigma_{\text{e}}^2$.}
  \label{fig:antennas}
\end{figure}

Fig.~\ref{fig:Symbolerrorrate} illustrates the impact of the number of antennas. The symbol error rate decreases when the number of transmit and receive antennas increases, which demonstrates the benefits of larger MIMO configurations. Fig.~\ref{fig:antennas}(a) shows the weighting mechanism on the user-intent relevance loss (BLIP-2) $\mathcal{L}_{\text{user}}^{\text{BLIP-2}}$ under different SNRs. Here, we introduce two additional baselines, \textit{QO-ISC-Random} and \textit{QO-ISC-PA}. QO-ISC-Random uses the same architecture as QO-ISC but all features have the random weight in SA-HBF. QO-ISC-PA determines hybrid beamforming matrices using the conventional MMSE HBF algorithm~\cite{Lin2019} and exploits the semantic importance weights to allocate the transmit power across subcarriers in the digital precoding. Since per-subcarrier power allocation is infeasible under constraint~\eqref{eq:P0_constraint1}, we relax constraint~\eqref{eq:P0_constraint1} to a total power constraint of $KP$ for QO-ISC-PA. Its total transmit power for all subcarriers is the same as that of the proposed SA-HBF. Among the weighting mechanisms under constraint~\eqref{eq:P0_constraint1}, the user-intent relevance loss $\mathcal{L}_{\text{user}}^{\text{BLIP-2}}$ of QO-ISC consistently outperforms the QO-ISC using uniform weight or random weight. This highlights the effectiveness of the proposed contextual-bandit weighting mechanism in SA-HBF, which prioritizes query-relevant features and mitigates semantic distortion during transmission. When $N_{\text{t}}=N_{\text{r}}=64$ and the SNR $=-25$ dB, the proposed QO-ISC achieves the user-intent relevance loss $\mathcal{L}_{\text{user}}^{\text{BLIP-2}}$ of 0.19, while QO-ISC-PA achieves 0.185. This is because the learned weights are concentrated on a small number of subcarriers, and power allocation directly increases the received SNR of the prioritized subcarriers, whereas the analog beamformer is shared by all subcarriers and its weighted steering provides a smaller gain. This result still shows that the learned semantic weights are effective under both the beamforming and the power allocation mechanisms.
Fig.~\ref{fig:antennas}(b) shows the impact of imperfect CSI on the proposed SA-HBF. The estimated channel is modeled by the statistical CSI error model, given by $\hat{\mathbf{H}}[k] = \mathbf{H}[k] + \mathbf{E}[k]$, where $\mathbf{E}[k]$ denotes the channel estimation error matrix with $\operatorname{vec}(\mathbf{E}[k])\sim\mathcal{CN}\left(\mathbf{0},\sigma_{\text{e}}^2 \mathbf{I}_{N_{\text{t}}N_{\text{r}}}\right)$ and $\sigma_{\text{e}}^{2}$ denotes the variance of the channel estimation error. The hybrid beamforming matrices are computed based on the estimated channel, while the transmission is performed over the true channel. The performance of the proposed QO-ISC degrades gracefully as $\sigma_{\text{e}}^2$ increases. The degradation is small at SNR $=-25$ dB, since the noise dominates the CSI estimation error. The degradation is more noticeable at SNR $=-10$ dB, since the beamforming mismatch caused by the estimation error dominates the performance. Moreover, QO-ISC achieves a lower $\mathcal{L}_{\text{user}}^{\text{BLIP-2}}$ than QO-ISC-Uniform for all evaluated values of $\sigma_{\text{e}}^2$. This shows that the gain of the proposed semantic-aware weighting is preserved under imperfect CSI, and the results under perfect CSI represent upper-bound results.

\subsection{Impact of User-intent Relevance Loss}
\begin{figure}[t]
    \centering
    \begin{subfigure}[b]{0.24\textwidth}
        \centering
        \includegraphics[width=\textwidth]{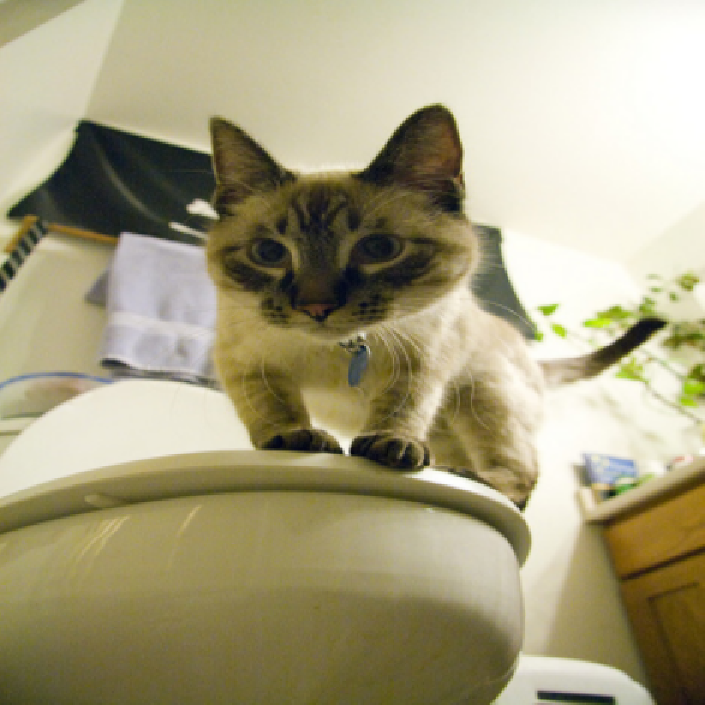}
        \caption{}
        \label{fig:subb}
    \end{subfigure}
    \begin{subfigure}[b]{0.24\textwidth}
        \centering
        \includegraphics[width=\textwidth]{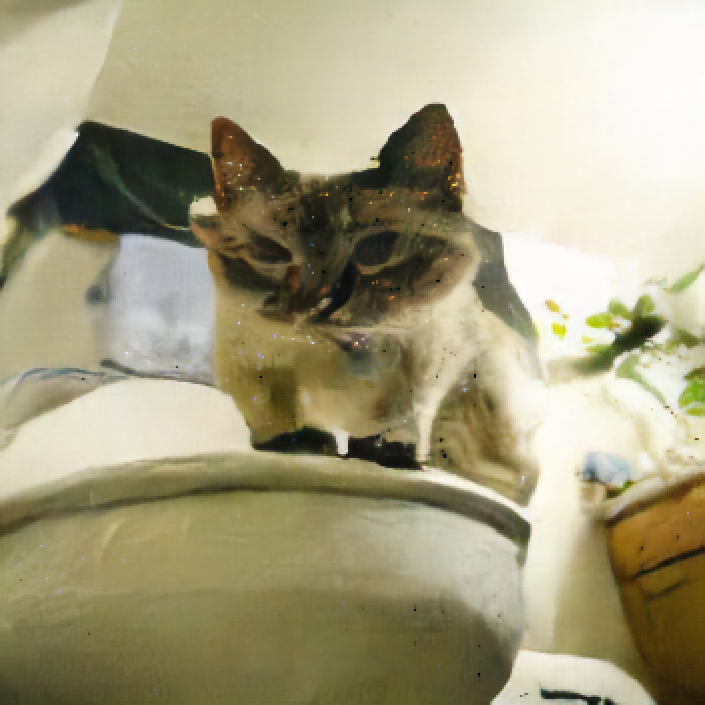}
        \caption{}
        \label{fig:subb}
    \end{subfigure}
    \begin{subfigure}[b]{0.24\textwidth}
        \centering
        \includegraphics[width=\textwidth]{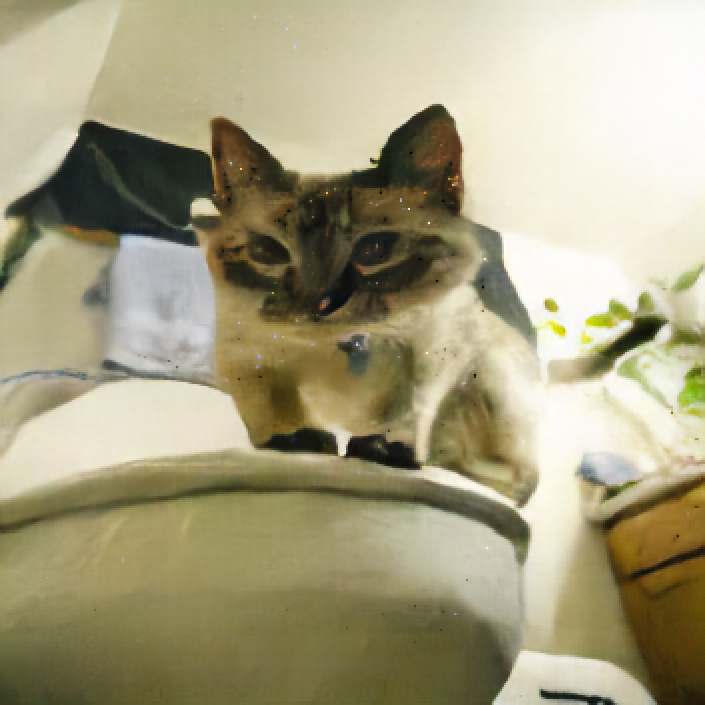}
        \caption{}
        \label{fig:subb}
    \end{subfigure}
    \begin{subfigure}[b]{0.24\textwidth}
        \centering
        \includegraphics[width=\textwidth]{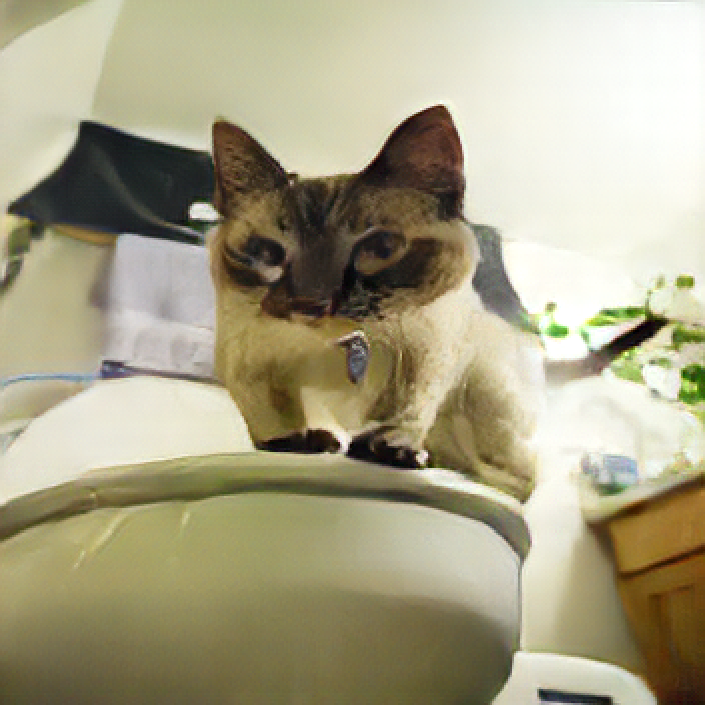}
        \caption{}
        \label{fig:subb}
    \end{subfigure}
    \caption{(a) Visualization of the original image. Visualization of the reconstructed images for (b) $\lambda_{\text{user}} = 0.0$, (c) $\lambda_{\text{user}} = 1.0$, and (d) $\lambda_{\text{user}} = 1.5$, under the query ``Does the cat have a collar?'' at SNR $= -22$~dB with $\lambda_{\text{gen}} = 0.5$.}
    \label{fig:visual-lambda-user}
\end{figure}
The parameter $\lambda_{\text{user}}$ in equation~(\ref{eq:L_total}) is a weighting coefficient used during the training phase to control the tradeoff between the user-intent relevance loss and other objectives, such as the $\ell_1$ loss and the adversarial loss. Specifically, the user-intent relevance loss focuses on preserving local semantic details aligned with the query, whereas other loss functions such as $\ell_1$ and adversarial losses aim to maintain the global structural fidelity and perceptual realism of the reconstructed image. Increasing the value of $\lambda_{\text{user}}$ encourages the model to prioritize the specific regions related to the user intent. In Fig.~\ref{fig:visual-lambda-user}, the user query is ``Does the cat have a collar?", making the cat's collar the query-relevant semantic element. We set $\lambda_{\text{gen}}$ to be 0.5 in these experiments and vary $\lambda_{\text{user}}$ from 0.0 to 1.5. When $\lambda_{\text{user}} = 0$ in Fig.~\ref{fig:visual-lambda-user}(b), the user-intent relevance loss (LLaVA) is disabled in training and the query-specific features are not prioritized, causing the collar details to be very blurred. As $\lambda_{\text{user}}$ increases to $1.0$ in Fig.~\ref{fig:visual-lambda-user}(c), the collar details become more apparent as the model places more emphasis on the query relevant region. When $\lambda_{\text{user}} = 1.5$ in Fig.~\ref{fig:visual-lambda-user}(d), the model further prioritizes the collar region, leading to better reconstruction of the query-relevant semantic content. 
To quantify this behavior, we conduct simulations and determine $\mathcal{L}_{\text{user}}^{\text{BLIP-2}}$ using different values $\lambda_{\text{user}}$. As shown in Fig.~\ref{fig:lambda_user_loss_weight}, $\mathcal{L}_{\text{user}}^{\text{BLIP-2}}$ decreases monotonically as $\lambda_{\text{user}}$ increases from $0.0$ to $1.5$ under different SNR conditions, confirming that a larger $\lambda_{\text{user}}$ steers the model toward prioritizing the user's intent. Furthermore, the effect of $\lambda_{\text{user}}$ is more significant at SNR $= -22$~dB than at SNR $= -10$~dB. This is because at low SNR, the corresponding lower channel capacity limits the amount of information that can be reliably delivered to the receiver. Under this constraint, the training objective with a larger $\lambda_{\text{user}}$ provides higher incentive for the encoder to concentrate its coding capacity on query-relevant semantic content. At high SNR, the channel capacity is sufficient to provide reliable transmission of both query-relevant and query-irrelevant features, so the reconstruction quality becomes less sensitive to $\lambda_{\text{user}}$.
\begin{figure}[t]
    \centering
    \includegraphics[width=0.6\linewidth]{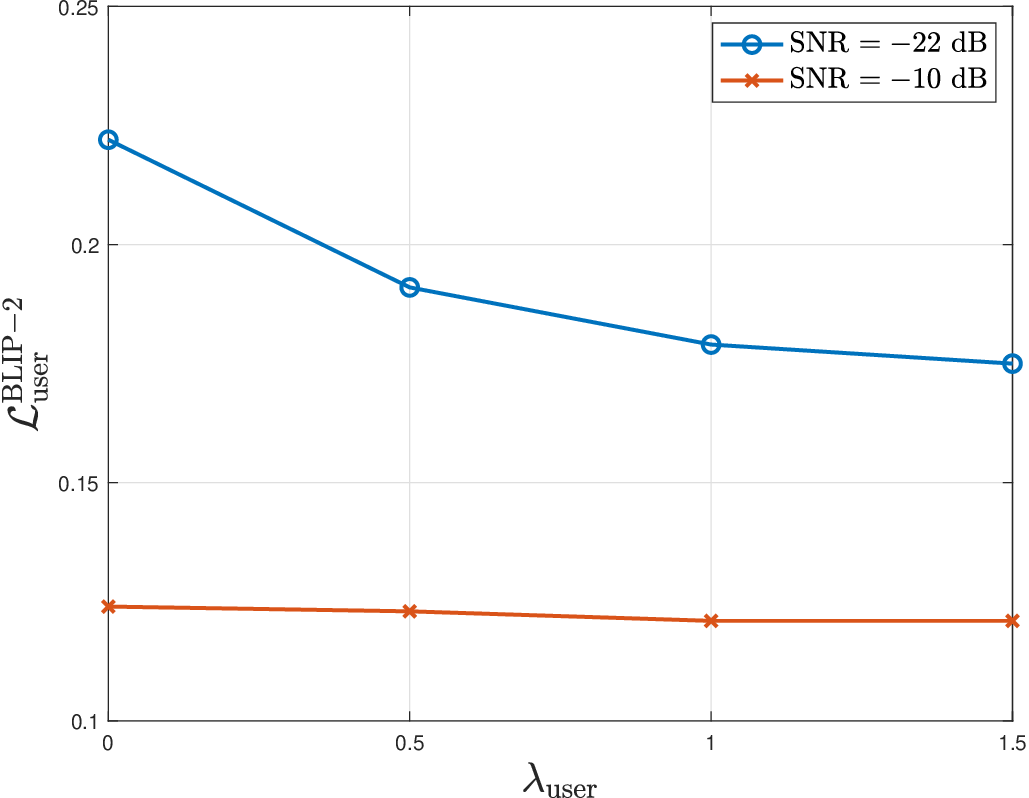}
    \caption{The user-intent relevance loss (BLIP-2) $\mathcal{L}_{\text{user}}^{\text{BLIP-2}}$ versus the tunable parameter $\lambda_{\text{user}}$.}
    \label{fig:lambda_user_loss_weight}
\end{figure}
\subsection{Impact of Query Type}
\begin{figure}[t]
    \centering
    \includegraphics[width=0.7\linewidth]{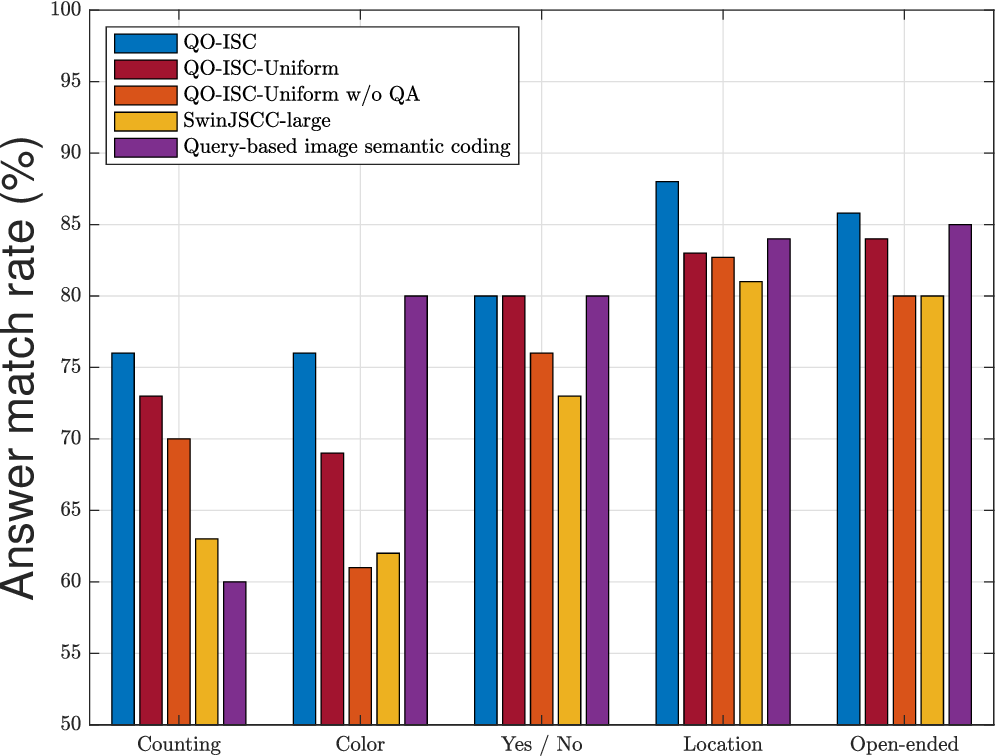}
    \caption{The answer match rate versus different types of questions.}
    \label{fig:Accuracybytype}
\end{figure}
Fig.~\ref{fig:Accuracybytype} presents the answer match rate across five question types for the baseline semantic coding schemes. The SNR is set to $-10$ dB. Our proposed QO-ISC achieves the highest performance across most categories. Notably, QO-ISC has the highest answer rate in the ``Location" and ``Open-ended" questions. Removing the weighting mechanism slightly degrades the performance across all the questions, with a particularly noticeable drop in ``Color" questions. The query-based semantic coding baseline achieves competitive results in ``Color" questions since the provided answer helps guide the diffusion model, but falls short in ``Counting" questions. Overall, QO-ISC can effectively handle different question types.

\subsection{Impact of Adversarial Training}
\begin{figure}[t]
  \centering
    
   \begin{subfigure}[b]{0.235\textwidth}
      \centering
      \includegraphics[width=\linewidth]{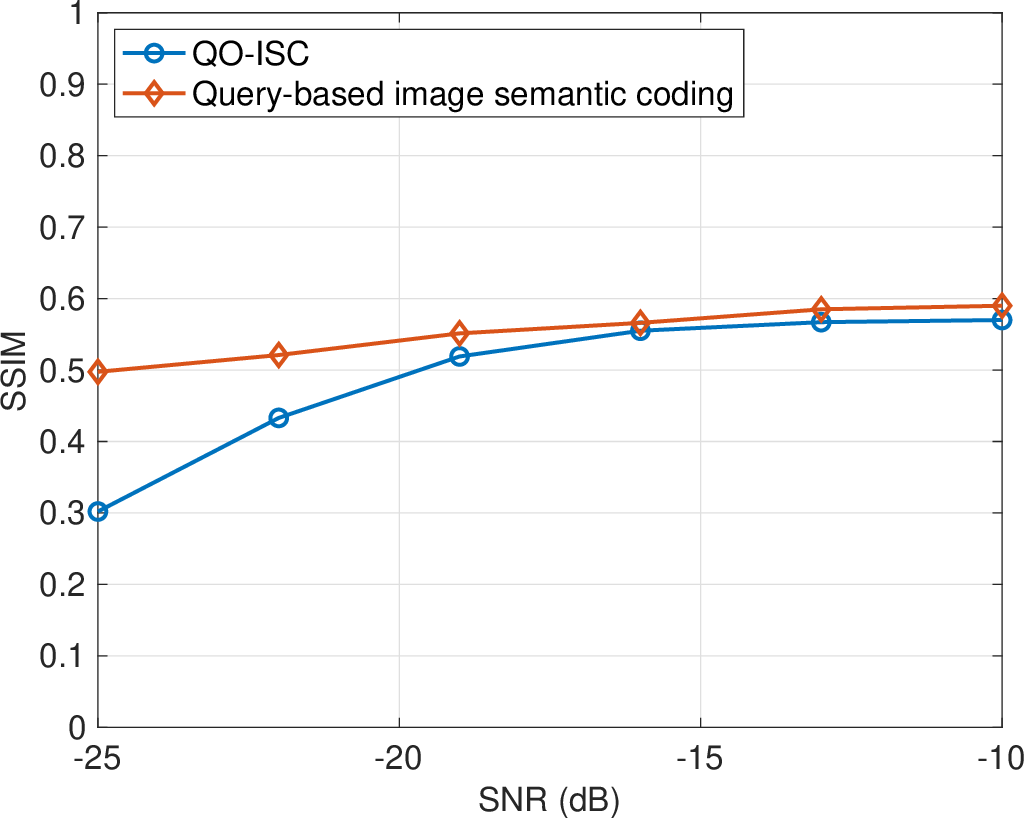}
       \caption{}
   \end{subfigure}
   \hfill
   \begin{subfigure}[b]{0.235\textwidth}
       \centering
       \includegraphics[width=\linewidth]{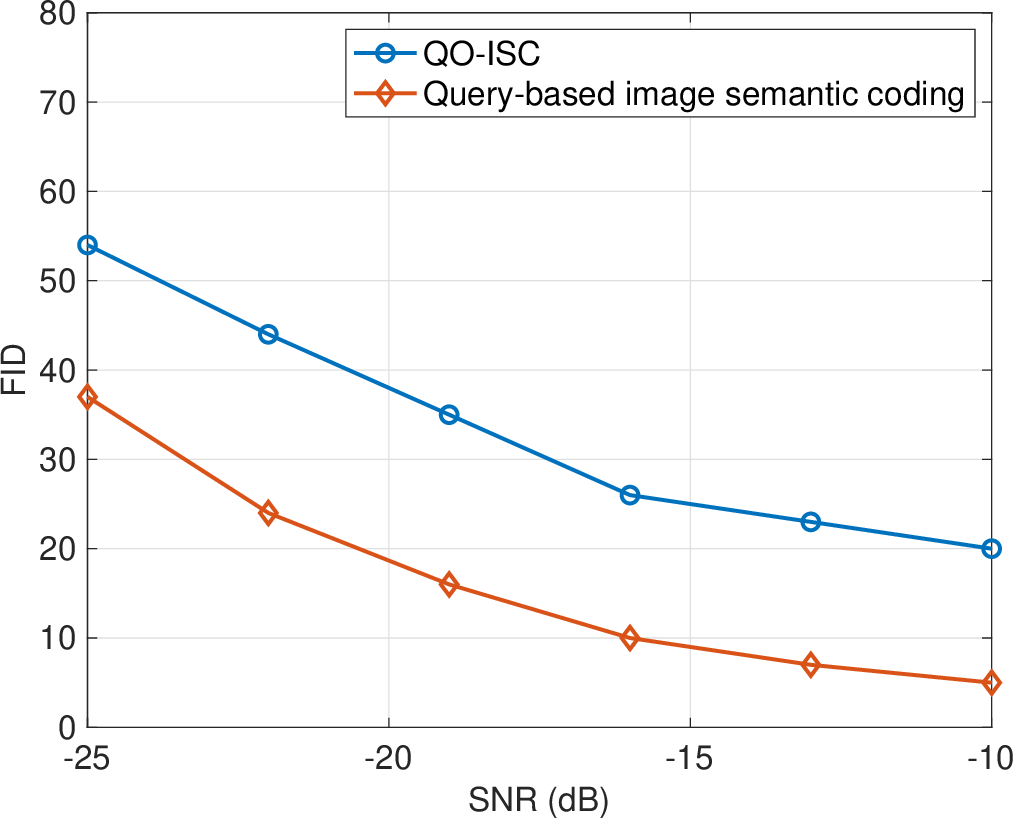}
       \caption{}
   \end{subfigure}
  \caption{(a) SSIM and (b) FID versus the SNR. Higher SSIM indicates better performance, while lower FID indicates better performance.}
  \label{fig:adv}
\end{figure}

We use the structural similarity index measure (SSIM) and Fréchet inception distance (FID) to evaluate the visual quality of the reconstructed images, as shown in Figs.~\ref{fig:adv}(a) and (b), respectively. SSIM measures the structural fidelity relative to the ground truth, while FID evaluates the semantic similarity by comparing the feature distributions between real and generated images. In these figures, we only compare our proposed QO-ISC against query-based image semantic coding, since SwinJSCC suffers from generalization issues and JPEG2000 fails to reconstruct images at low SNR levels. Since the query-based image semantic coding baseline uses a diffusion model to iteratively refine the output, it achieves a higher SSIM and a lower FID compared to QO-ISC. This performance gap stems from the pretrained knowledge of the diffusion model, which leverages the iterative denoise process to create high-fidelity textures and sharp details. However, as demonstrated in our previous results, although the baseline produces more general reconstructions, our proposed QO-ISC provides a higher answer match rate, indicating that our framework can effectively preserve the specific semantic features required to fulfill the user's intent.

\subsection{Impact of Vector Quantization}
Recall from Section~\ref{Sec:system} that we partitioned each embedding vector into multiple segments and quantized each segment using a set of codewords. Fig.~\ref{fig:quantization} illustrates the impact of the feature vector segment length $N_{L}$ on the user-intent relevance loss (BLIP-2) $\mathcal{L}_{\text{user}}^{\text{BLIP-2}}$ under different SNR levels. A shorter segment length divides the feature vector into smaller segments, resulting in lower user-intent relevance loss $\mathcal{L}_{\text{user}}^{\text{BLIP-2}}$ and more robust performance, since changes in a single segment affect only part of the feature vector. Although shorter segments can reduce $\mathcal{L}_{\text{user}}^{\text{BLIP-2}}$, they increase the number of transmitted symbols. Specifically, each embedding vector of dimension $D$ is partitioned into $L = D/N_L$ segments. Since each segment is mapped to a specific codeword index for transmission, the total number of transmitted symbols is directly proportional to $L$. This creates a trade-off where decreasing $N_L$ leads to a higher number of segments $L$. Thus, it enhances the performance of the system. Although this finer partitioning can reduce $\mathcal{L}_{\text{user}}^{\text{BLIP-2}}$, it increases the communication overhead. 
\begin{figure}[t]
    \centering
    \includegraphics[width=0.7\linewidth]{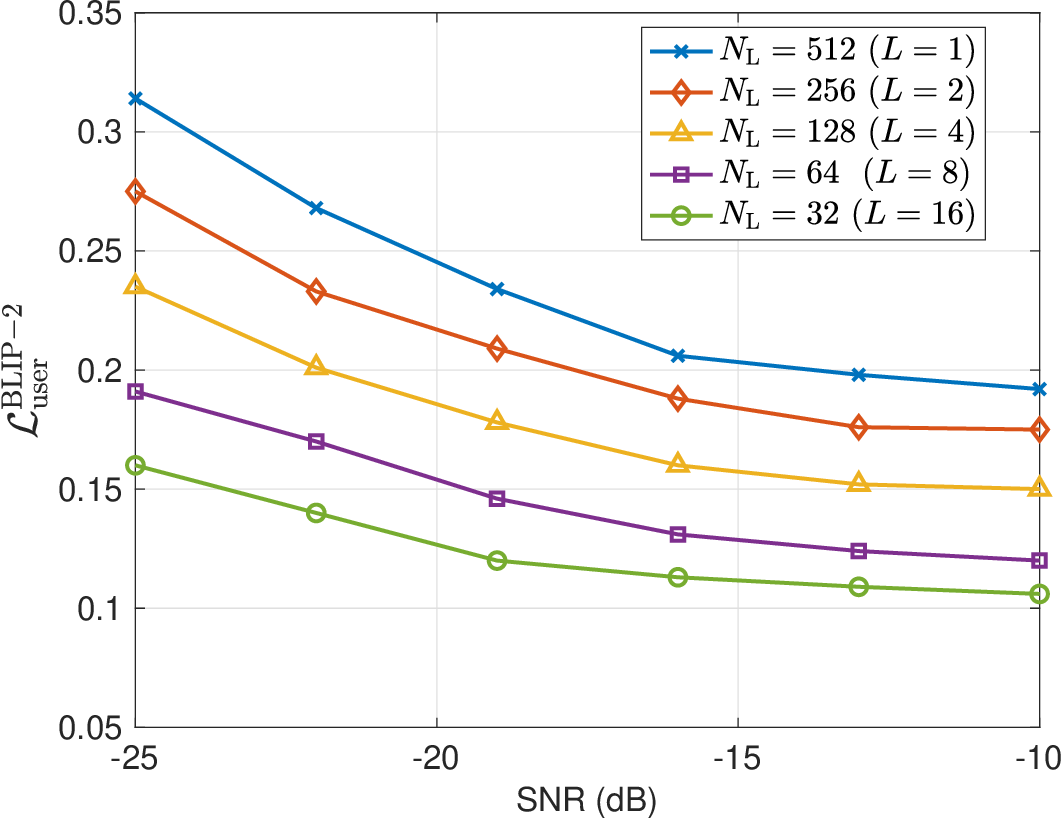}
    \caption{The user-intent relevance loss (BLIP-2) $\mathcal{L}_{\text{user}}^{\text{BLIP-2}}$ versus the SNR under different feature vector segment length $N_{\text{L}}$.}
    \label{fig:quantization}
\end{figure}

\subsection{Analysis of the Number of Transmitted Symbols and Model Computational Complexity}
\begin{table}[t]
    \centering
    \caption{\small\textsc{Comparison of Transmitted Symbols, 
Model Parameters, and Computational Complexity Per Image}}
    \label{tab:symbols}
    \begin{tabular}{|c|c|c|c|c|}
        \hline
        \textbf{Models} & \makecell{QO-ISC}
        & \makecell{Query-based \\ image semantic \\ coding}
        & \makecell{SwinJSCC- \\ large}
        & \makecell{JPEG \\ 2000} \\
        \hline
         \makecell{\textbf{Number of} \\ \textbf{Transmitted}\\ \textbf{Symbols}} & 61,952 & 15,860 & 495,616 & 82,603 \\
        \hline
         \makecell{\textbf{Number of}\\ \textbf{Total}\\ \textbf{Parameters}} & 165.5M & 9.1B & 52.1M & N/A \\
        \hline
         \makecell{\textbf{Number of} \\ \textbf{Trainable}\\ \textbf{Parameters}} & 15.88M & 722M & 52.07M & N/A \\
        \hline
        \makecell{\textbf{GFLOPs}} & 115.8 & 4,945 & 319.5 & N/A \\
        \hline
        \end{tabular}
\end{table}
As shown in Table~\ref{tab:symbols}, we compare the number of transmitted symbols, number of model parameters, and computational complexity of QO-ISC against the baselines. The proposed QO-ISC transmits only 61,952 symbols per image, which is 25\% fewer than JPEG2000 and $8\times$ fewer than SwinJSCC-large. This reduction is achieved because SwinJSCC relies on floating-point feature transmission, whereas QO-ISC employs discrete vector quantization. The query-based image semantic coding~\cite{chen2024} has the lowest number of  transmitted symbols because it transmits a highly compressed image along with a text description. However, it relies on a diffusion-based decoder to reconstruct the image from the text description, which can lead to hallucinations as shown in the visualization results in Fig.~\ref{fig:visual}. Regarding the model size, although QO-ISC has 165M parameters, most of these parameters belong to the frozen CLIP backbone and only 9.6\% of the total parameters are trainable. Furthermore, despite having more total parameters than SwinJSCC-large, QO-ISC requires fewer GFLOPs because CLIP ViT-B/16 processes only 484 tokens per image using a patch size of 16, compared to 30,976 spatial positions processed by SwinJSCC using a patch size of 2. The query-based image semantic coding baseline incurs the highest GFLOPs due to its reliance on a 7B-parameter LLM and a 50-step diffusion decoder. In summary, QO-ISC achieves the balance across all three dimensions. It transmits fewer symbols than both JPEG2000 and SwinJSCC, requires fewer GFLOPs than SwinJSCC despite more total parameters, and limits its training overhead to only 15.88M trainable parameters through a frozen CLIP backbone. 

We further report the online computational overhead. The semantic encoder, weight module, and decoder are executed on an NVIDIA T4 GPU, with average runtimes of 10.1 ms, 0.1 ms, and 4.3 ms per image, respectively. The WMMSE hybrid beamforming algorithm is executed on two AMD EPYC 7302 CPUs with a runtime of approximately 3 sec per image. This step is the bottleneck of the online processing, because each iteration involves matrix inversions and the manifold optimization. Since the weight matrix depends on the transmitted features, the hybrid beamforming matrices are recomputed for each image and query pair.




\section{conclusion}
\label{Sec:conclude}
In this paper, we proposed a generalized QO-ISC framework for large-scale MIMO-OFDM system. To enhance generalization, we incorporated a pretrained LAM to extract the image and text features. To capture user's intent, we aligned the image features with text features for image reconstruction. To prioritize semantically important features in a large-scale MIMO-OFDM system, we developed the SA-HBF algorithm. Zero-shot experiments demonstrated that our proposed generalized QO-ISC framework outperforms the state-of-the-art JSCC codec, SwinJSCC, by 7.6\% in terms of the answer match rate by using a pretrained LAM backbone. In low SNR region, incorporating text alignment provides a 9\% gain, while employing importance weighting in the SA-HBF improves the performance  by 4.8\%. Our QO-ISC framework also outperforms the state-of-the-art query-based image semantic coding method by 8\%. For the future work, we plan to extend the proposed framework to multiuser scenario, which is more practical for deployment.

\appendix
\section{WMMSE Hybrid beamforming Algorithm}
\label{app:wmmse}
 Following the formulation in~\cite{Lin2019}, the digital precoder $\mathbf{V}_{\text{D}}[k]$ can be expressed as $\beta_k \mathbf{V}_{\text{U}}[k]$, where $\mathbf{V}_{\text{U}}[k]$ denotes the digital precoding matrix without normalization. Substituting \eqref{eq:received_signal} into the MSE term in \eqref{eq:P2a}, the WMMSE hybrid precoding problem can be formulated as:
\begin{subequations} \label{eq:hbf_problem}
    \begin{align}
        \underset{\mathbf{V}_{\mathrm{A}},\, \mathbf{V}_{\mathrm{U}},\, \boldsymbol{\beta}}{\text{minimize}} \quad &
        \sum_{k=1}^{K} w_{b,k} \, \mathrm{tr} \Big( 
        \mathbf{\dot{H}}^H[k] \mathbf{V}_{\mathrm{A}} \mathbf{V}_{\mathrm{U}}[k] \mathbf{V}_{\mathrm{U}}^H [k]
        \mathbf{V}_{\mathrm{A}}^H \mathbf{\dot{H}}[k] \nonumber \\
        &\quad - \mathbf{\dot{H}}^H[k] \mathbf{V}_{\mathrm{A}} \mathbf{V}_{\mathrm{U}}[k] 
        - \mathbf{V}_{\mathrm{U}}^H[k] \mathbf{V}_{\mathrm{A}}^H \mathbf{\dot{H}}[k] \nonumber \\
        &\quad + \sigma^2 \beta_k^{-2} \mathbf{Q}^H[k] \mathbf{Q}[k] 
        + \mathbf{I}_{N_s} \Big) \label{eq:hbf_problem_obj} \\
        \text{subject to} \quad &
        \mathrm{tr} \left( \mathbf{V}_{\mathrm{A}} \mathbf{V}_{\mathrm{U}}[k] 
        \mathbf{V}_{\mathrm{U}}^H[k] \mathbf{V}_{\mathrm{A}}^H \right) \leq \beta_k^{-2}P, 
        ~ k \in \mathcal{K}, \label{eq:hbf_problem_power} \\
        & \text{constraint }~\eqref{eq:P0_constraint2}, \nonumber
    \end{align}
\end{subequations}
where $\mathbf{\dot{H}}[k]=\mathbf{H}^{H}[k]\mathbf{Q}_{\text{A}}\mathbf{Q}_{\text{D}}[k]$ is the effective channel matrix that incorporates the fixed combiners. For a fixed analog precoder $\mathbf{V}_{\mathrm{A}}$, the power is fully utilized when $\beta_{k}=\sqrt{P}\left(\mathrm{tr}(\mathbf{V}_{\text{A}}\mathbf{V}_{\text{U}}[k]\mathbf{V}_{\text{U}}^{H}[k]\mathbf{V}_{\text{A}}^H)\right)^{-1/2}$. By applying the Karush-Kuhn-Tucker (KKT) conditions, the closed-form solution of the digital precoder $\mathbf{V}_{\text{U}}[k]$ can be derived as
\begin{align}
    \mathbf{V}_{\text{U}}[k]~ 
    =~&w_{b,k} \Big( 
        \mathbf{V}_{\mathrm{A}}^H \, \dot{\mathbf{H}}[k] \, 
        \dot{\mathbf{H}}^H[k] \, \mathbf{V}_{\text{A}} \nonumber \\
        &+ \frac{\sigma^2}{P} \mathrm{tr}(\mathbf{Q}^{H}[k]\mathbf{Q}[k]) \, 
        \mathbf{V}_{\text{A}}^H \, \mathbf{V}_{\text{A}} 
    \Big)^{-1} 
    \mathbf{V}_{\text{A}}^H \, \dot{\mathbf{H}}[k].
    \label{eq:VDk}
\end{align}
Substituting the optimal $\mathbf{V}_{\text{U}}[k]$ and $\beta_k$ into the MSE term in \eqref{eq:hbf_problem_obj}, the MSE for subcarrier $k$ can be expressed as:
\begin{align}
    \mathcal{J}_{k}(\mathbf{V}_{\text{A}}) \triangleq 
    \mathrm{tr} \bigg( \Big( \mathbf{I}_{N_s} 
    &+ \frac{P}{\sigma^2} \mathrm{tr}(\mathbf{Q}^{H}[k]\mathbf{Q}[k]) \mathbf{\dot{H}}^H[k] \mathbf{V}_{\text{A}}
     \nonumber \\
    &\times \left( \mathbf{V}_{\text{A}}^H \mathbf{V}_{\text{A}} \right)^{-1}\mathbf{V}_{\text{A}}^H \mathbf{\dot{H}}[k] \Big)^{-1} \bigg).
    \label{eq:JVRF}
\end{align}
The optimization of the analog precoder $\mathbf{V}_{\text{A}}$ corresponds to the following WMMSE problem: 
\begin{align}
    \underset{\mathbf{V}_{\text{A}}}{\text{minimize}} \quad &
    \sum_{k=1}^{K} w_{b,k} \, \mathcal{J}_{k}(\mathbf{V}_{\mathrm{A}}) \label{eq:optVRF_obj} \\
    \text{subject to } & \text{constraint}~\eqref{eq:P0_constraint2}. \nonumber
\end{align} 
Since each entry of the analog precoder matrix $\mathbf{V}_{\text{A}}$ lies on the complex unit circle, problem \eqref{eq:optVRF_obj} can be solved by the manifold optimization (MO) approach. The Euclidean gradient is projected onto the tangent space and is updated using the Armijo-Goldstein line search. The iterate is retracted back to the manifold to preserve the modulus constraint.

After optimizing the hybrid precoder, we proceed to optimize the hybrid digital and analog combiners $\mathbf{Q}_{\text{D}}[k]$ and $\mathbf{Q}_{\text{A}}$ with the fixed precoding matrices and normalizing factor $\beta_{k}$. Similar to the precoding problem, we include the weight vector $\mathbf{w}_{b}$ from the contextual bandit into the hybrid combining algorithm. Substituting \eqref{eq:received_signal} into the MSE term in~\eqref{eq:P2a}, the hybrid combining problem can be formulated as:
\begin{equation}\label{eq:hbf_combiner_obj}
\begin{aligned}
    \underset{\mathbf{Q}_{\text{A}},\, \mathbf{Q}_{\text{D}}}{\text{minimize}} \quad &
    \sum_{k=1}^{K} w_{b,k} \, \mathrm{tr} \Big( 
    \mathbf{Q}^{H}[k]\mathbf{ \ddot{H}}[k] \mathbf{\ddot{H}}^{H}[k] \mathbf{Q}[k] \\
    &\quad - \mathbf{Q}^{H}[k] \mathbf{\ddot{H}}[k] 
    - \mathbf{\ddot{H}}^{H}[k] \mathbf{Q}[k] \\
    &\quad + \sigma^2 \beta_k^{-2} \mathbf{Q}^{H}[k] \mathbf{Q}[k] 
    + \mathbf{I}_{N_s} \Big) \\
    \text{subject } & \text{to constraint}~\eqref{eq:P0_constraint3},
\end{aligned}
\end{equation}
where $\mathbf{\ddot{H}}[k] = \mathbf{H}[k] \mathbf{V}_{\text{A}} \mathbf{V}_{\text{D}}[k]$ is the effective channel matrix that incorporates the fixed precoders. The analog combiner is fixed. By differentiating the weighted MSE function~\eqref{eq:hbf_combiner_obj} with respect to $\mathbf{Q}_{\text{D}}[k]$, we derive the optimal $\mathbf{Q}_{\text{D}}[k]$ as:
\begin{align}
    \mathbf{Q}_{\text{D}}[k] 
    = w_{b,k} \Big( 
        &\mathbf{Q}_{\mathrm{A}}^H \, \ddot{\mathbf{H}}[k] \, 
        \ddot{\mathbf{H}}^H[k] \, \mathbf{Q}_{\text{A}} \nonumber \\
        &+ \sigma^2 \beta_{k}^{-2}\mathbf{Q}_{\text{A}}^{H}\mathbf{Q}_{\text{A}} \, 
        \mathbf{V}_{\text{A}}^H \, \mathbf{V}_{\text{A}} 
    \Big)^{-1} 
    \mathbf{V}_{\text{A}}^H \, \ddot{\mathbf{H}}[k].
    \label{eq:QDk}
\end{align}
By substituting the optimal $\mathbf{Q}_{\text{D}}[k]$ into the objective function~\eqref{eq:hbf_combiner_obj}, we can express the MSE as follows:
\begin{align}
    \mathcal{G}_{k}(\mathbf{Q}_{\text{A}}) \triangleq 
    \mathrm{tr} \bigg( \Big( \mathbf{I}_{N_s} 
    &+ \frac{1}{\sigma^2 \beta_{k}^{-2}} \mathbf{\ddot{H}}^H[k] \mathbf{Q}_{\text{A}}
    \left( \mathbf{Q}_{\text{A}}^H \mathbf{Q}_{\text{A}} \right)^{-1} \nonumber \\
    &\times \mathbf{Q}_{\text{A}}^H \mathbf{\ddot{H}}[k] \Big)^{-1} \bigg).
    \label{eq:JARF}
\end{align}
The optimization of the analog combiner $\mathbf{Q}_{\text{A}}$ corresponds to the following WMMSE problem:
\begin{align}
        \underset{\mathbf{Q}_{\text{A}}}{\text{minimize}} \quad &
        \sum_{k=1}^{K} w_{b,k} \, \mathcal{G}_{k}(\mathbf{Q}_{\mathrm{A}}) \label{eq:opt_QA_obj} \\
        \text{subject to } & \text{constraint}~\eqref{eq:P0_constraint3}. \nonumber
\end{align}   
Problem \eqref{eq:opt_QA_obj} is similar to the analog precoding problem \eqref{eq:optVRF_obj}. Therefore, we can apply the same MO-based algorithm to obtain $\mathbf{Q}_{\mathrm{A}}$. By alternately solving the precoding and combining subproblems, we iteratively update the beamforming matrices to minimize the semantic-aware weighted MSE. The WMMSE hybrid beamforming algorithm is shown in Algorithm~\ref{alg:mmseHBF}.

\begin{algorithm}[t]
\small
\caption{\small WMMSE Hybrid Beamforming Algorithm}
\label{alg:mmseHBF}
\begin{algorithmic}[1]
\State \textbf{Input:} The channel response $\mathbf{H}$;  the power limit $P$; the noise variance $\sigma^2$; the weight vector $\mathbf{w}_{b}$ for transmission block $b$; the number of iterations $N_{\text{HBF}}$.
\State Randomly initialize hybrid precoder matrices $\mathbf{V}_{\text{A}}$, $\mathbf{V}_{\text{D}}$, and hybrid combiner matrices $\mathbf{Q}_{\text{A}}$, $\mathbf{Q}_{\text{D}}$.
\For{$i \gets 1$ \textbf{to} $N_{\text{HBF}}$} 
    \State \textbf{I. Optimize Hybrid Precoder}
    \State Keep the hybrid combiner matrices $\mathbf{Q}_{\text{A}}$ and $\mathbf{Q}_{\text{D}}$ to be fixed.
    \State Update the unnormalized digital precoder $\mathbf{V}_{\text{U}}[k]$ by~\eqref{eq:VDk} and normalizing factor $\beta_{k}:=\sqrt{P}\left(\mathrm{tr}(\mathbf{V}_{\text{A}}\mathbf{V}_{\text{U}}[k]\mathbf{V}_{\text{U}}^{H}[k]\mathbf{V}_{\text{A}}^H)\right)^{-1/2}$ for $k = 1, \ldots, K$.
    \State Solve the analog precoder $\mathbf{V}_{\text{A}}$ problem in~\eqref{eq:optVRF_obj} via MO.
    \State Update the digital precoder $\mathbf{V}_{\text{D}}[k] := \beta_k \mathbf{V}_{\text{U}}[k]$.
    \State \textbf{II. Optimize Hybrid Combiner}
    \State Keep the hybrid precoder matrices $\mathbf{V}_{\text{A}}$ and $\mathbf{V}_{\text{D}}$ to be fixed.
    \State Update the digital combiner $\mathbf{Q}_{\text{D}}[k]$ by~\eqref{eq:QDk} for $k \in \mathcal{K}$.
    \State Solve the analog combiner $\mathbf{Q}_{\text{A}}$ problem in~\eqref{eq:opt_QA_obj} via MO.
\EndFor
\State \textbf{Output}: The optimized hybrid beamforming matrices $\{\mathbf{V}_{\text{A}}^{*}, \mathbf{V}_{\text{D}}^{*}, \mathbf{Q}_{\text{A}}^{*}, \mathbf{Q}_{\text{D}}^{*}\}$.
\end{algorithmic}
\end{algorithm}

\bibliographystyle{ieeetr}
\bibliography{refs}

\begin{IEEEbiography}[{\includegraphics[width=1in,height=1.25in,clip,keepaspectratio]{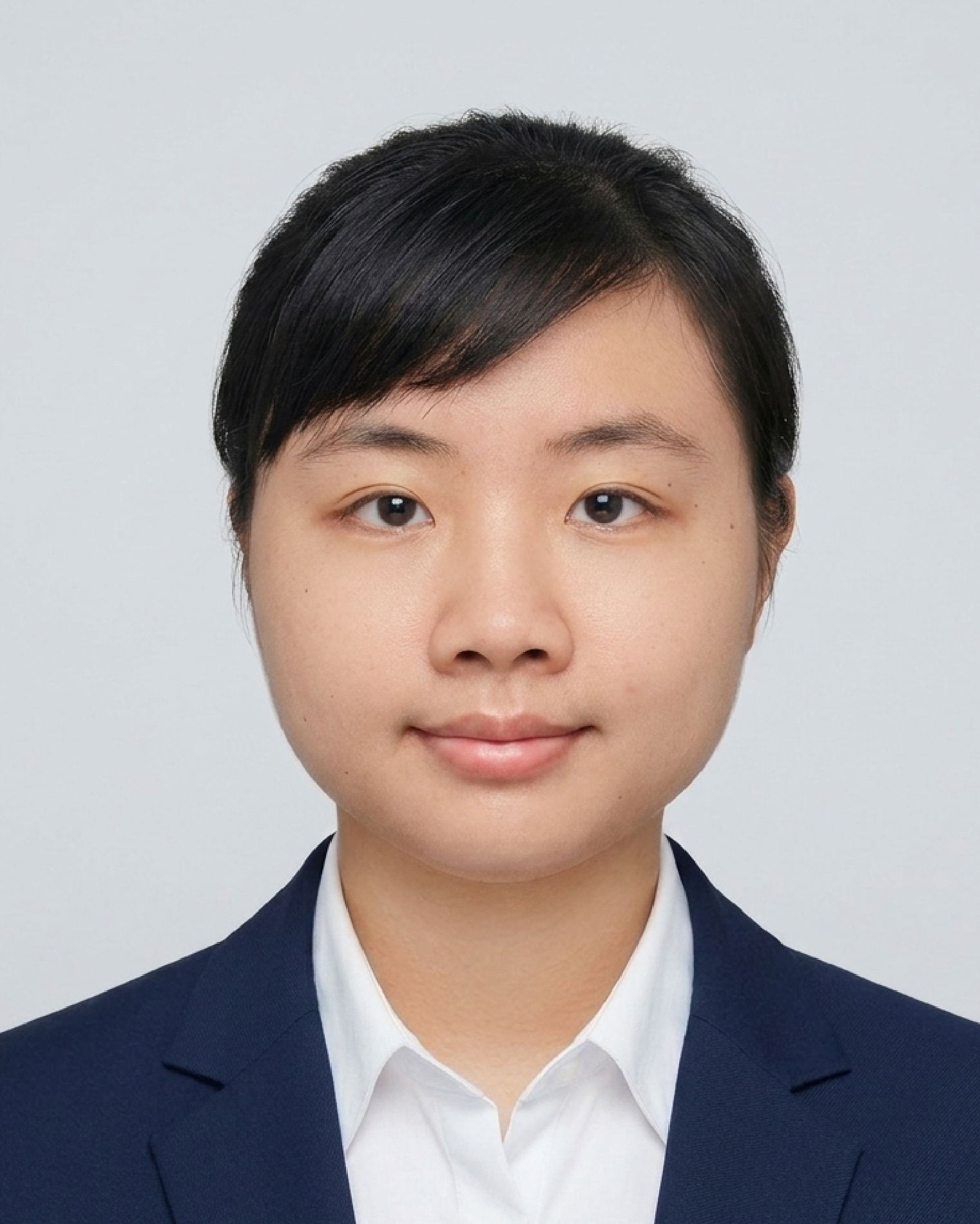}}]{Sin-Yu\;Huang} (Graduate Student Member, IEEE) received the B.S. degree in electrical engineering and the M.S. degree in computer and communication engineering from National Cheng Kung University, Tainan, Taiwan, in 2021 and 2022, respectively. From 2022 to 2023, she was a Research Assistant with the Research Center for Information Technology Innovation, Academia Sinica, Taipei, Taiwan. She is currently pursuing the Ph.D. degree with the Department of Electrical and Computer Engineering, The University of British Columbia (UBC), Vancouver, BC, Canada. Her research interests include semantic communications, task-oriented communications, large AI models for wireless networks, and agentic AI for 6G systems. She is a recipient of the President's Academic Excellence Initiative Ph.D. Award from UBC. She has served as a TPC Member and Session Chair of the \textit{IEEE ICC} 2026.
\end{IEEEbiography}

\begin{IEEEbiography}[{\includegraphics[width=1in,height=1.25in,clip,keepaspectratio]{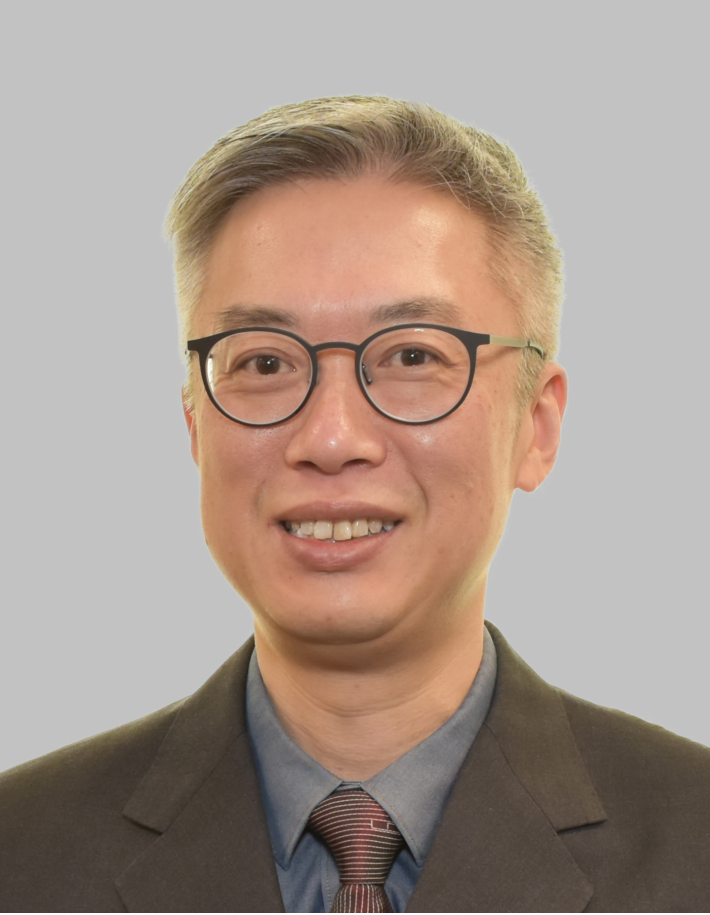}}]{Vincent\;W.S.\;Wong} (S'94, M'00, SM'07, F'16) received the B.Sc. degree from the University of Manitoba, Canada, in 1994, the M.A.Sc. degree from the University of Waterloo, Canada, in 1996, and the Ph.D. degree from the University of British Columbia (UBC), Vancouver, Canada, in 2000. From 2000 to 2001, he worked as a systems engineer at PMC-Sierra Inc. (now Microchip Technology Inc.). He joined the Department of Electrical and Computer Engineering at UBC in 2002 and is currently a Professor and UBC Distinguished University Scholar. His research areas include protocol design, optimization, and resource management of communication networks, with applications to 5G/6G wireless networks, Internet of things, mobile edge computing, smart grid, and energy systems. Dr. Wong is the Editor-in-Chief of the {\it IEEE Transactions on Wireless Communications}. He has served as an Area Editor of the {\it IEEE Transactions on Communications} and {\it IEEE Open Journal of the Communications Society}, an Associate Editor of the {\it IEEE Transactions on Mobile Computing} and {\it IEEE Transactions on Vehicular Technology}, and a Guest Editor of the {\it IEEE Journal on Selected Areas in Communications}, {\it IEEE Internet of Things Journal}, and {\it IEEE Wireless Communications}. Dr. Wong is the TPC Vice-Chair of {\it IEEE GLOBECOM} 2027, Publicity Co-Chair of {\it IEEE VTC2027-Spring}, and Area Chair of {\it IEEE INFOCOM} 2027. He has served as the General Co-Chair of {\it IEEE INFOCOM} 2024; Tutorial Co-Chair of {\it IEEE GLOBECOM} 2018; Technical Program Co-Chair of {\it IEEE  VTC}2020{\it -Fall} and {\it IEEE SmartGridComm} 2014; and Symposium Co-Chair of {\it IEEE ICC}'18, {\it IEEE SmartGridComm} ('13, '17) and {\it IEEE GLOBECOM}'13. He received the 2022 Best Paper Award from {\it IEEE Transactions on Mobile Computing} and Best Paper Awards at the {\it IEEE ICC} 2022 and {\it IEEE GLOBECOM} 2020. He has served as the Chair of the IEEE Vancouver Joint Communications Chapter and IEEE Communications Society Emerging Technical Sub-Committee on Smart Grid Communications. He was an IEEE Communications Society Distinguished Lecturer from 2019 to 2020 and is an IEEE Vehicular Technology Society Distinguished Lecturer for the term of 2023$-$2026. Dr. Wong is a Fellow of the IEEE, Canadian Academy of Engineering, and the Engineering Institute of Canada.
\end{IEEEbiography}

\end{document}